\documentclass[11pt, a4paper]{article}
\usepackage[margin=25mm]{geometry}
\usepackage{lmodern}
\usepackage{amsthm}
\usepackage{amsmath}
\usepackage{amsfonts}
\usepackage[shortlabels]{enumitem}
\usepackage{booktabs}
\usepackage{multirow}
\usepackage{amssymb}%
\usepackage{mathrsfs}
\usepackage{textcomp}
\usepackage{authblk}
\usepackage{graphicx}
\usepackage[numbers,compress]{natbib}
\usepackage{mathtools}
\usepackage[colorlinks=true, linkcolor=blue, urlcolor=blue, citecolor=blue]{hyperref}
\usepackage{bm}
\usepackage{bbm}
\usepackage{epstopdf}
\usepackage[caption=false]{subfig}
\newcounter{subeq}

\begin{document}
\title{A generalized fundamental solution technique for the regularized 13-moment-system in rarefied gas flows}
\author[1]{Himanshi\thanks{Corresponding author: \href{hkhungar@gmail.com}{\tt{hkhungar@gmail.com}},  \href{phd2001141003@iiti.ac.in}{\tt{phd2001141003@iiti.ac.in}}}}
\author[2]{Lambert Theisen}
\author[3]{Anirudh Singh Rana}
\author[2]{Manuel Torrilhon}
\author[1]{Vinay Kumar Gupta}
\affil[1]{Department of Mathematics, Indian Institute of Technology Indore,\newline Indore 453552, India}
\affil[2]{Applied and Computational Mathematics, RWTH Aachen University, \newline Aachen 52062, Germany}
\affil[3]{Department of Mathematics, Birla Institute of Technology and Science Pilani, Rajasthan 333031, India}
%

\date{}
\maketitle
\thispagestyle{empty}

\begin{abstract}
In this work, we explore the method of fundamental solutions (MFS) for solving the regularized 13-moment (R13) equations for rarefied monatomic gases.
While previous applications of the MFS in rarefied gas flows relied on problem-specific fundamental solutions, we propose a generic approach that systematically computes the fundamental solutions for any linear moment system without predefined source terms.
The generalized framework is first introduced using a simple example involving the Stokes equations, and is then extended to the R13 equations.
The results obtained from the generic MFS are validated against an analytical solution for the R13 equations.
Following validation, the framework is applied to the case of thermally-induced flow between two non-coaxial cylinders.
Since no analytical solution exists for this case, we compare the results obtained from the MFS with those obtained from the finite element method (FEM). To further assess computational efficiency, we analyze the runtimes of the FEM and MFS.
The results indicate that the MFS converges faster than the FEM and serves as a promising alternative to conventional meshing-based techniques.
\end{abstract}

\section{\label{sec:intro}Introduction}
Advancements in micro- and nano-machining have led to the miniaturization of mechanical and electrical devices, especially in systems like micro-heat exchangers, pumps, turbines, and sensors.
The design and operation of these devices rely on understanding how gases flow and transfer heat in rarefied conditions.
 Traditional fluid models assume gases have very short mean-free-paths which makes them inaccurate for rarefied gases, where molecules travel longer distances before colliding.
Classical fluid dynamics equations, like the Euler or Navier--Stokes--Fourier (NSF) equations, are based on the assumption of near-equilibrium and very short mean free paths. Such assumptions break down in rarefied environments and lead to significant inaccuracies in predicting flow fields. To address this limitation, extended hydrodynamic models have been formulated by incorporating additional moments of the distribution function. 

The regularized 13-moment (R13) equations~\citep{ST2003,Struchtrup2005}  represent one such extended hydrodynamic model, which includes additional evolution equations for the stress tensor and heat flux vector to offer a decent approximation of the Boltzmann equation.
The R13 equations predict the presence of Knudsen layers, allow for smooth shock structures, and are third-order accurate in the Knudsen number.
 A variety of numerical techniques have been applied to solve the R13 equations, including finite difference schemes, discontinuous Galerkin approaches, and finite element approaches~\citep{RTS2013,TS2017,
 WT2019,TT2021}.
 While these mesh-based methods are effective, they require significant computational resources, especially for complex geometries where mesh generation can be time-consuming and computationally expensive.

An alternative approach to mesh-based methods is the MFS, a mesh-free method that offers significant advantages over traditional numerical techniques.
Originally introduced in the 1960s~\citep{KA1964}, applications of the MFS have outspread several fields in past few decades, including electromagnetics, elasticity, and fluid mechanics~\citep{BK2001,FKM2003,
YJFMT2006}.
The MFS approximates the solution of boundary value problems by representing the solution as a linear combination of the fundamental solutions.
 These fundamental solutions are the exact solutions to the governing differential equation with singularities at specified source points located outside the problem domain.
 The coefficients of these fundamental solutions are determined by enforcing boundary conditions at discrete points on the boundary of the domain.
 The absence of the meshing of the domain makes the MFS particularly suitable for problems involving complex and evolving geometries, such as shape optimization and inverse problems.
  Additionally, the MFS avoids numerical integration which sets it apart from other mesh-free methods, like the boundary element method.

  Given these advantages, there has been growing interest in applying the MFS to rarefied gas flows~\citep{LC2016,CSRSL2017,
  RSCLS2021,HRG2023,HRG2025}.
   Traditionally, the classical MFS constructs the solution as a linear combination of fundamental solutions, where unknown coefficients represent the strengths of point sources.
   These coefficients are determined by enforcing the boundary conditions.
    An alternative perspective emerged by drawing inspiration from the Stokeslet---the fundamental solution to the Stokes equations.
    In this viewpoint, the source strengths are introduced directly into the governing equations as multipliers of Dirac delta functions before deriving the fundamental solution. As a result, the derived solutions inherently incorporate the physical meaning of the sources, such as point forces.
 Building on this idea, Lockerby and Collyer~\citep{LC2016} proposed a method in which the classical coefficients are replaced by physically interpretable point forces and point heat sources embedded within the momentum and energy balance equations.
  These unknown strengths are then determined using boundary conditions.
  The formulation in~\citep{LC2016} led to the derivation of fundamental solutions in three dimensions and their implementation for the Grad 13-moment (G13) equations~\cite{Grad1949}.
  In Ref.~\citep{CSRSL2017}, an additional (ad hoc) source term was introduced in the stress evolution equation to obtain fundamental solutions of the R13 equations in three dimensions.
   Apart from that, in order to obtain the fundamental solution for the CCR model~\citep{RGS2018}, \citet{RSCLS2021} and~\citet{HRG2023} simply used a source term in the mass balance equation in addition to the momentum and energy balance equations for investigating evaporation effects in three and two dimensions, respectively.
   All these approaches required deriving fundamental solutions for specific models by prescribing Dirac-delta source terms in selected governing equations within the system and/or in the closure relations.
    While effective, this methodology typically makes it challenging to extend the MFS for new or more complex models, where the fundamental solutions are unknown and the choice of source terms is not straightforward.
    In these formulations, the number of unknown source terms at each singularity point can be interpreted as the degrees of freedom associated with that point.
    These degrees of freedom determine how the fundamental solutions contribute to the overall solution.

    To address the limitations posed by fixing the source terms manually, we propose a generic approach that allows for the computation of fundamental solutions for any large system of equations without the need to predefine specific Dirac-delta source terms.
The generic MFS approach relies on two steps.
The first step involves identifying the fundamental solutions of the system.
This process draws inspiration from Hörmander's method~\cite{Hormander1955,Banerjee1994} and employs Fourier transformation in combination with partial fraction decomposition to derive expressions for the fundamental solutions.
The second step is determining the source strengths using the boundary conditions for the problem under consideration.

Before applying this method to a complex system of equations, we first demonstrate its implementation on a simple example---Stokes' equations.
Ultimately, we extend the generic MFS to the R13 equations.
We implement and derive the fundamental solutions of the R13 equations in two dimensions and validate it against an analytical solution.
After validation, we consider a problem of thermally induced flow between two noncoaxial cylinders for which analytic solution is not known.
Therefore, the results obtained from the MFS are compared with those obtained from the finite element method (FEM).
The ability of the MFS to efficiently capture rarefaction effects without requiring domain meshing makes it an attractive alternative to traditional numerical methods, such as the FEM.
 Additionally, we provide a comparative analysis of the computational effort required for implementing the MFS and FEM in order to highlight the advantages of the proposed approach in terms of efficiency and accuracy.

Despite its many advantages, the accuracy of the MFS is known to be highly sensitive to the placement of singularities or source points\citep{Alves2009, CKL2016, WLQ2018, CH2020}.
The optimal location of these sources depends on the grid spacing or the number of boundary nodes and source points.
Another common challenge with implementing MFS is the ill-conditioning of the collocation matrix.
As observed in previous studies~\cite{Alves2009,CH2020}, there exists a trade-off between accuracy and conditioning---improving one often leads to the deterioration of the other.
To address this, several studies~\cite{DML2009,CNYC2023,WL2011,HRG2023,HRG2025} have used the effective condition number, which provides a more reliable indicator than the traditional condition number, to optimize the placement of singularities for improved accuracy in the MFS.
In this work, we also examine the impact of the singularity locations, the grid spacing, and the effective condition number on the accuracy of the MFS.

The rest of the paper is structured as follows.
The R13 equations in linear and steady state with thermodynamically consistent boundary conditions are described in Sec.~\ref{sec:r13model}.
A brief of the generic MFS approach applicable to any two-dimensional linear moment system is given in Sec.~\ref{sec:approach}.
The approach has been elaborated for a simple example of Stokes' equations in Sec.~\ref{sec:stokes}.
The core implementation of the generic MFS for the R13 equations is detailed in Sec.~\ref{sec:r13_mfs}.
The validation of the results obtained from the MFS with the analytic solution for the R13 equations in the problem of two coaxial cylinders is given in Sec.~\ref{sec:analytic}.
The comparison between the MFS and FEM for thermally induced flow between two noncoaxial cylinders is presented in Sec.~\ref{sec:fem} followed by a conclusion and outlook in Sec.~\ref{sec:conclusion}.

\section{\label{sec:r13model}The R13 Equations and Boundary Conditions}
This section introduces the steady-state and linearized R13 equations, since the MFS relies upon the linearity of the equations.
The density $\tilde \rho$ and classical temperature $\tilde T$ are expressed with pressure $\tilde p$ and temperature $\tilde\theta$ as done in~\cite{TS2017,TT2021}---the full R13 equations can be found in~\cite{TorrilhonARFM, Struchtrup2005}.
The collision frequency $\tilde\nu$ appearing in the right-hand sides of the balance equations for heat flux $\tilde{\bm{q}}$ and stress $\tilde{\bm{\sigma}}$ is reformulated using $\tilde p = \tilde \rho \tilde \theta$ and $\tilde u = 3 \tilde\theta / 2$ for monatomic ideal gases.
 To nondimensionalize and linearize the equations, perturbations in flow variables from their respective equilibrium states are considered.
 The reference equilibrium density and temperature are $\tilde \rho_0$ and $\tilde \theta_{0}$, whereas the velocity, stress and heat flux vanish in the equilibrium state.
 All the variables with tilde denote the dimensional quantities while those without tilde are dimensionless.
Considering $\tilde{L}$ as the physical length scale, the dimensionless position vector $\bm{x}: \Omega\to \mathbb{R}^3$, temperature $\theta: \Omega\to \mathbb{R}$, pressure $p: \Omega\to \mathbb{R}$ and velocity $\bm{v}: \Omega\to \mathbb{R}^3$ read
\begin{align}
\bm{x} = \frac{\tilde{\bm{x}}}{\tilde L}, \quad \theta = \frac{\tilde \theta}{\tilde \theta_0}, \quad p = \frac{\tilde p}{\tilde p_0}, \quad \bm{v} = \frac{\tilde{\bm{v}}}{\sqrt{\tilde \theta_0}},
\end{align}
respectively, and other dimensionless quantities are
\begin{align}
\bm{\sigma} = \frac{\tilde{\bm{\sigma}}}{\tilde p_0}, \quad \bm{q} = \frac{\tilde{\bm{q}}}{\tilde p_0 \sqrt{\tilde \theta_0}}, \quad \bm{m} = \frac{\tilde{\bm{m}}}{\tilde p_0 \sqrt{\tilde\theta_0}}, \quad \bm{R} = \frac{\tilde{\bm{R}}}{\tilde p_0 \tilde \theta_0},
\quad \triangle = \frac{\tilde \triangle}{\tilde p_0 \tilde \theta_0}.
\end{align}
Here, $\bm{\sigma}: \Omega\to \mathbb{R}^{3\times 3}$ and $\bm{R}: \Omega\to \mathbb{R}^{3\times 3}$ are symmetric trace-free second-order tensors, while $\bm{m}: \Omega\to \mathbb{R}^{3\times 3 \times 3}$ is a symmetric trace-free third-order tensor.
The resulting system of linear, steady state, and dimensionless R13 equations read
\begin{align}
\label{mass_bal}
    \bm{\nabla} \cdot \bm{v} &= 0, \\ \label{mom_bal}
    \bm{\nabla} p + \bm{\nabla} \cdot \bm{\sigma} &= \bm{0},
    \\ \label{energy_bal}
    \bm{\nabla} \cdot \bm{q} &= 0,
    \\
 \label{sigma_bal}
    \frac{4}{5} \overline{\bm{\nabla} \bm{q}} + 2 \overline{\bm{\nabla} \bm{v}} + \bm{\nabla} \cdot \bm{m} &= -\frac{1}{\text{Kn}} \bm{\sigma},
    \\
    \label{q_bal}
    \frac{5}{2} \bm{\nabla} \theta + \bm{\nabla} \cdot \bm{\sigma} + \frac{1}{2} \bm{\nabla} \cdot \bm{R} + \frac{1}{6} \bm{\nabla} \triangle &= -\frac{1}{\text{Kn}} \frac{2}{3} \bm{q},
\end{align}
with the closure
\begin{align}
 \label{R_bal}
    \bm{R} &= -\frac{24}{5} \text{Kn} \overline{\bm{\nabla} \bm{q}},
    \\
\label{m_bal}
    \bm{m} &= -2 \text{Kn} \overline{\bm{\nabla} \bm{\sigma}},
    \\ \label{delta_bal}
    \triangle &= -12 \text{Kn} \bm{\nabla} \cdot \bm{q}.
\end{align}
Here, $\mathrm{Kn}=\tilde{\tau}_0\sqrt{\tilde{\theta}_0}/\tilde{L}$ is the Knudsen number, with $\tilde{\tau}=1/\tilde{\nu}$ being the mean free time.
The overline in the terms denotes the symmetric and trace-free (deviatoric) part of the tensors.
The symmetric trace-free part of a rank-$2$ tensor $\bm{J} \in \mathbb{R}^{3 \times 3}$ is defined component-wise as
\begin{align}
\overline{\bm{J}} = J_{\langle ij\rangle} = J_{(ij)} - \frac{1}{3} J_{kk} \delta_{ij} = \frac{1}{2} \left(J_{ij} + J_{ji}\right) - \frac{1}{3} J_{kk} \delta_{ij},
\end{align}
where $\delta_{ij}$ is the Kronecker's delta function.
For instance, the symmetric trace-free part of rank-2 tensor $\bm{\nabla}\bm{v}$ reads
\begin{align}
    \overline{\bm{\nabla}\bm{v}} = \frac{\partial v_{\langle i}}{\partial x_{j\rangle}}=\frac{1}{2}\left(\frac{\partial v_i}{\partial x_j}+\frac{\partial v_j}{\partial x_i}\right) - \frac{1}{3} \delta_{ij} \frac{\partial v_k}{\partial x_k}.
\end{align}
The symmetric and trace-free part of a rank-3 tensor $\bm{K} \in \mathbb{R}^{3 \times 3 \times 3}$ analogously reads
\begin{align}
\overline{\bm{K}} = K_{\langle ijk\rangle} = K_{(ijk)} - \frac{1}{5} \left(K_{(ill)} \delta_{jk} + K_{(ljl)} \delta_{ik} + K_{(llk)} \delta_{ij}\right),
\end{align}
where
\begin{align}
     K_{(ijk)}=\frac{1}{6}\left( K_{ijk}+K_{ikj}+K_{jik}+K_{jki}+K_{kij}+K_{kji} \right)
\end{align}
denotes the symmetric part of tensor $\bm{K}$. For example, the symmetric trace-free part of tensor $\bm{\nabla}\bm{\sigma}$ is
\begin{align}
\overline{\bm{\nabla}\bm{\sigma}} = \frac{\partial \sigma_{\langle i j}}{\partial x_{k\rangle}} = \frac{1}{3} \left(
\frac{\partial \sigma_{ij}}{\partial x_k}
+ \frac{\partial \sigma_{ik}}{\partial x_j}
+ \frac{\partial \sigma_{jk}}{\partial x_i}
\right)
- \frac{2}{15} \left(
\frac{\partial \sigma_{km}}{\partial x_m} \delta_{ij}
+ \frac{\partial \sigma_{jm}}{\partial x_m} \delta_{ik}
+ \frac{\partial \sigma_{im}}{\partial x_m} \delta_{jk}
\right).
\end{align}
Utilizing~\eqref{energy_bal} in~\eqref{delta_bal}, we obtain $\triangle=0$.

Throughout this work, problem domains $\hat{\Omega} \subset \mathbb{R}^3$ are considered, which are homogeneous along the $z$-direction, representing cross-sections of infinitely extended domains.
This allows the reduction of the problem to a two-dimensional (2D) computational domain $\Omega \subset \mathbb{R}^2$ by assuming $\partial_z \equiv 0$.
While this assumption simplifies the tensor variables, the velocity space and tensor structure formally remain three-dimensional (3D).
The only independent components corresponding to this assumption are given by the vector
{\small
\setlength\arraycolsep{3pt}
\begin{align*}
\bm{U} = \begin{bmatrix}
p & v_x & v_y & \sigma_{xx} & \sigma_{xy} & \sigma_{yy} & \theta & q_x & q_y & m_{xxx} & m_{xxy} & m_{yyx} & m_{yyy} & R_{xx} & R_{xy} & R_{yy}
\end{bmatrix}^\mathsf{T}.
\end{align*}
}

%
The thermodynamically admissible linearized 2D boundary conditions for the R13 equations are
\begin{align}
    \label{bc_vn}
(\bm{v}-\bm{v}^{\mathrm{w}})\cdot\bm{n} &=     \epsilon^{\mathrm{w}} \tilde{\chi} \left( p - p^{\mathrm{w}} + \bm{n}\cdot\bm{\sigma}\cdot\bm{n}  \right)
     \\ \label{bc_sigma_nt}
   \bm{n}\cdot\bm{\sigma}\cdot\bm{t}  &= \tilde{\chi} \left(\bm{v}-\bm{v}^{\mathrm{w}} + \frac{1}{5}\bm{q}   + \bm{n}\cdot\bm{m}\cdot\bm{n} \right)\cdot\bm{t},
   \\ \label{bc_R_nt}
   \bm{n}\cdot\bm{R}\cdot\bm{t} &= \tilde{\chi} \left( -(\bm{v}-\bm{v}^{\mathrm{w}}) + \frac{11}{5}\bm{q} -\bm{n}\cdot\bm{m}\cdot\bm{n}\right)\cdot\bm{t},
    \\ \label{bc_q_n}
   \bm{q} \cdot\bm{n} &= \tilde{\chi} \left( 2(\theta - \theta^{\mathrm{w}}) + \frac{1}{2}\bm{n}\cdot\bm{\sigma}\cdot\bm{n}  + \frac{2}{5}\bm{n}\cdot\bm{R}\cdot\bm{n}  \right),
     \\ \label{bc_m_nnn}
    (\bm{n}\cdot\bm{m}\cdot\bm{n})\cdot\bm{n} &= \tilde{\chi} \left( -\frac{2}{5}(\theta - \theta^{\mathrm{w}}) + \frac{7}{5}\bm{n}\cdot\bm{\sigma}\cdot\bm{n} - \frac{2}{25}\bm{n}\cdot\bm{R}\cdot\bm{n}   \right),
     \\ \label{bc_mntt}
   \bm{n}\cdot\left( \frac{1}{2}\bm{n}\cdot\bm{m}\cdot\bm{n} + \bm{t}\cdot\bm{m}\cdot\bm{t}\right) &= \tilde{\chi} \left( \frac{1}{2}\bm{n}\cdot\bm{\sigma}\cdot\bm{n} + \bm{t}\cdot\bm{\sigma}\cdot\bm{t} \right),
\end{align}
where $\bm{n}$ and $\bm{t}$ are the unit normal and tangent vectors.
Further, $\tilde{\chi}$ denotes the modified accommodation factor which is given by
\begin{align}
\tilde{\chi} = \frac{\sqrt{2/(\pi \theta_0)}\chi}{2-\chi}.
\end{align}
In Eq.~\eqref{bc_vn}, $\epsilon^{\mathrm{w}}$ is the velocity prescription coefficient used to implement artificial in- and outflow conditions with interface pressure $p^{\mathrm{w}}$ and velocity $\bm{v}^{\mathrm{w}}$.
This boundary condition is reduced to the standard boundary condition $\bm{v}\cdot \bm{n}=0$ for  $\bm{v}^{\mathrm{w}}=0$ and $\epsilon^{\mathrm{w}}=0$.
The dot product of two symmetric tensors $T_{i_1 i_2 \dots i_n}$ and $W_{j_1 j_2 \dots j_m}$ is defined as $\sum_{k} T_{i_1 i_2 \dots i_{n-1} k} W_{k j_2 \dots j_m} $. For instance, the dot products in~\eqref{bc_vn}--\eqref{bc_mntt} are $\bm{v}\cdot\bm{n}=\sum_i v_i n_i$,  $\bm{n}\cdot\bm{\sigma}\cdot\bm{t} = \sum_{i,j} \sigma_{ij}n_i t_j$ and $(\bm{n}\cdot\bm{m}\cdot\bm{n})\cdot\bm{n} = \sum_{i,j,k} m_{ijk}n_i n_j n_k$, and analogously for other tensors.

\section{\label{sec:approach}Implementing Generic Method of Fundamental Solutions}
This section introduces a general technique to determine and implement the fundamental solutions for any linear first-order system of partial differential equations.
We consider a linearized system of $N\in \mathbb{N}$ partial differential equations in (2D) Cartesian coordinates, expressed as
\begin{align}
\label{system}
\bm{A}^{(x)}\partial_x \bm{U} + \bm{A}^{(y)}\partial_y \bm{U} + \bm{P} \bm{U} = \bm{S}\delta(\bm{r}),
\end{align}
where $\bm{U}:\Omega\to \mathbb{R}^N$ is the variable vector field,  $\bm{A}^{(x)}, \bm{A}^{(y)} \in \mathbb{R}^{N\times N}$ are constant advection matrices and $\bm{P} \in \mathbb{R}^{N\times N}$ is the constant reaction matrix, $\bm{S}\in\mathbb{R}^N$ is a constant forcing vector (including source terms) and $\delta(\bm{r})$ is the Dirac delta.
To determine the fundamental solution of the system, we define the Fourier transform $\hat{F}(\bm{k})$ of a function $F(\bm{r})$ as
\begin{align}
\label{ft}
\mathcal{F}\big(F(\bm{r})\big)
=\hat{F}(\bm{k}) := \int_{\mathbb{R}^2} F(\bm{r}) \, \mathrm{e}^{-\mathbbm{i} \, \bm{k} \cdot \bm{r}} \, \mathrm{d} \bm{r},
\end{align}
where $\mathbbm{i}$ is the imaginary unit, $\bm{k}=(k_x,k_y)\in \mathbb{R}^2$ is the wave vector in the spatial-frequency domain.
The corresponding inverse Fourier transformed counterpart is defined as
\begin{align}
\label{ftinv}
\mathcal{F}^{-1}\big(\hat{F}(\bm{k})\big)
=F(\bm{r})
:=\frac{1}{(2\pi)^2}\displaystyle{ \int_{\mathbb{R}^2}}
\hat{F}(\bm{k}) \, \mathrm{e}^{\mathbbm{i}\,\bm{k} \cdot \bm{r}} \, \mathrm{d}\bm{k}.
\end{align}
Applying Fourier transformation on Eq.~\eqref{system} we obtain
\begin{align}
\label{system_ft}
\bm{A}(\bm{k})\hat{\bm{U}} :=(\mathbbm{i} k_x \bm{A}^{(x)} + \mathbbm{i} k_y \bm{A}^{(y)}+ \bm{P})\hat{\bm{U}}=\bm{S} \hat{\delta},
\end{align}
 wherein the inverse of the matrix $\bm{A}\in \mathbb{R}^{N\times N}$ can be written as
\begin{align}
\bm{A}(\bm{k})^{-1}=\frac{1}{\mathrm{det}(\bm{A}(\bm{k}))} \bm{\mathcal{A}}(\bm{k})=\frac{1}{s(\bm{k})}\bm{\mathcal{A}}(\bm{k}).
\label{inverse_adjugate_symbol}
\end{align}
Here, the determinant $ \mathrm{det}(\bm{A}(\bm{k}))=s(\bm{k})$ is identified as the \emph{symbol}~\cite{Folland1995} of the partial differential operator and the matrix $\bm{\mathcal{A}}$ is the adjugate matrix, which contains the cofactor expansions of the original matrix.
 Since both adjugate matrix and symbol contain polynomial terms in $k_x$ and $k_y$, they can be easily inverted using the Fourier inverse transformation.
 Using the fact that
 \begin{align}
     \bm{A}(\bm{\nabla})\bm{\mathcal{A}(\bm{\nabla})}=s(\bm{\nabla}) I_N,
 \end{align}
$I_N$ is the $N \times N$ identity matrix, one can conclude
 \begin{align}
     s(\bm{k})\hat{\bm{U}}=\bm{\mathcal{A}}(\bm{k})\bm{S}\hat{\delta} \iff s(\bm{\nabla})[\bm{U}]=\bm{\mathcal{A}}[\delta]\bm{S}.
 \end{align}
  This crucial step makes this approach commendable.
 Finding the fundamental solution corresponding to only the symbol operator leads us to the fundamental solution for the complete system.
 The fundamental solution for the full system is given by
 \begin{align}
 \label{fundamental_U}
 \bm{U}(\bm{r})=\bm{\mathcal{A}}(\bm{\nabla})[\Phi](\bm{r}) \bm{S},
 \end{align}
 where $\Phi$ is the 2D fundamental solution associated with the symbol $s(\bm{\nabla})$ of the PDE, i.e.
 \begin{align}
 s(\bm{\nabla})[\Phi]=\delta.
 \end{align}
 It is straightforward to calculate $\Phi$ if the symbol turns out to be a differential operator with a known fundamental solution.
 Furthermore, if the symbol can be factorized into Laplace, polyharmonic and Helmholtz operators, the fundamental solution $\Phi$ can be calculated using partial fraction decomposition along with inverse Fourier transforms of the known factor operators.
It is important to note that a fundamental solution $\Phi$ is not unique. Different solutions can be obtained by adding the homogeneous solutions, which correspond to the null space of the operator. This non-uniqueness plays a crucial role in constructing tailored solutions for specific boundary conditions and physical scenarios.

\begin{figure}[!t]
    \centering
\includegraphics[scale=0.8]{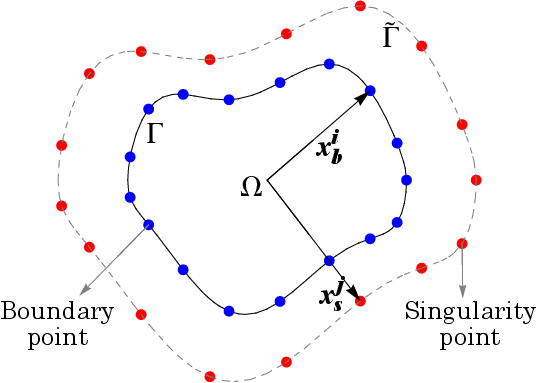}
    \caption{Schematic representation for discretization of boundary points (blue disks) on the domain boundary and singularity points (red disks) outside the problem domain.}%
    \label{fig:arb_schem}
\end{figure}

After finding the fundamental solution for the complete system, the MFS involves the discretization of the domain boundary into boundary nodes, also known as collocation points.
Furthermore, the singularity or source points are placed on some fictitious boundary outside the problem domain.
We demonstrate this by considering an arbitrary domain $\Omega$ having boundary $\Gamma$ as shown in Fig.~\ref{fig:arb_schem}.
The boundary $\Gamma$ is discretized into $n_{\text{b}}$ equispaced boundary points having position vectors $\bm{x}_j^{\text{b}}; \, j=1,\dots,n_{\text{b}}$.
Outside the domain $\Omega$, a fictitious boundary $\tilde\Gamma$ is considered with source points $\bm{x}_i^{\text{s}}; \, j=1,\dots,n_{\text{s}}$.
The relative position of the $i$th source point with respect to $j$th boundary node is denoted by $\bm{r}_{ij}=\bm{x}_j^{\text{b}}-\bm{x}_i^{\text{s}}$.
 The boundary conditions for the problem are written in the form
 \begin{align}
 \label{bn_U}
 \bm{B}(\bm{x}^{\text{b}}) \bm{U}(\bm{x}^{\text{b}})= \bm{g}(\bm{x}^{\text{b}}),
 \end{align}
 where $\bm{B}(\bm{x}^{\text{b}})\in \mathbb{R}^{p\times N}$ is a matrix depending on the normal and tangent vectors $\bm{n}$ and $\bm{t}$ associated with any point $\bm{x}^{\text{b}}$ lying on the boundary $\Gamma $ and $\bm{g}(\bm{x}^{\text{b}})\in \mathbb{R}^p$ is the corresponding right-hand-side vector.
The numerical solution obtained by the MFS is the superposition of the obtained fundamental solutions, i.e.
\begin{align}
    \bm{U}_{\text{MFS}} (\bm{x})=\sum_{i=1}^{n_{\text{s}}} \bm{\mathfrak{A}}(\bm{x}-\bm{x}_i^{\text{s}}) \bm{S}_i,
\end{align}
 where $\bm{U}_{\text{MFS}} (\bm{x})$ denotes the solution at any point $\bm{x}\in\Omega$,  $\bm{\mathfrak{A}}(\bm{r})\equiv \bm{\mathcal{A}}(\bm{\nabla})[\Phi](\bm{r})$ and $\bm{S}_i\in \mathbb{R}^{N}$ contains the unknown source strengths corresponding to $i$th source point $\bm{x}_i^{\text{s}}$.
 The unknown strengths are then calculated by solving a linear system formed on applying the boundary conditions at each boundary node.
 The linear system reads
 \begin{align}
     \bm{B}(\bm{x}_j^{\text{b}})\bm{U}(\bm{x}_j^{\text{b}})= \bm{B}(\bm{x}_j^{\text{b}})\sum_{i=1}^{n_{\text{s}}} \bm{\mathfrak{A}}(\bm{x}_j^{\text{b}}-\bm{x}_i^{\text{s}}) \bm{S}_i =\bm{g}(\bm{x}_j^{\text{b}}) , \quad j=1,2,\dots,n_{\text{b}}.
 \end{align}
 The overall linear system is $\bm{\mathcal{M}}\bm{X}=\bm{\mathcal{G}}$, where $\bm{\mathcal{M}}$ is the $p n_{\text{b}} \times N n_{\text{s}}$ collocation matrix, $\bm{X}\in\mathbb{R}^{N n_{\text{s}}}$ is the vector containing the unknown source strengths $\bm{S}_i$ corresponding to $i=1,2,\dots,n_{\text{s}}$ singularities and $\bm{\mathcal{G}}\in \mathbb{R}^{p n_{\text{b}}}$ contains the right-hand-side vectors $\bm{g}(\bm{x}_j^{\text{b}})$ for $j=1,2,\dots,n_{\text{b}}$.
Since the matrix $\bm{\mathcal{M}}$ is generally non-square, it is possible to have many equations ($N$) with comparatively fewer boundary conditions ($p$). The choice of the number of boundary and singularity points ($n_{\text{b}}$ and $n_{\text{s}}$, respectively) significantly influences the structure and solvability of the system. To facilitate a square system, we introduce a decomposition $\bm{S} = \bm{M} \bm{\mu}$, and choose $n_{\text{b}} = n_{\text{s}}$, so that the number of boundary conditions imposed at each boundary node matches the number of unknown source strengths associated with each singularity point.
This shall be discussed further in detail in the subsequent sections.

The approach elaborated above can be extended to 3D scenarios in a straightforward way by considering a 3D fundamental solution $\Phi$.

\section{\label{sec:stokes}Implementing the generic MFS for Stokes' equations}
 We show the implementation of generic MFS via an example of Stokes' equations (in two dimensions) which read
 \begin{align}
 \label{stokes1}
 \bm{\nabla}\cdot\bm{v}&=0,
 \\
 \label{stokes2}
 \bm{\nabla} p+ \bm{\nabla}\cdot\bm{\sigma}&=0,
 \\
  \label{stokes3}
\bm{\sigma}&=-\overline{\bm{\nabla}\bm{v}}.
 \end{align}
\subsection{\label{sec:fund_stokes}Fundamental solutions}
 Rewriting these equations as in the form of Eq.~\eqref{system}, the unknowns are \\ $\bm{U}=\big[
 p \;\; v_x \;\; v_y \;\; \sigma_{xx} \;\; \sigma_{xy} \;\; \sigma_{yy}
 \big]^\mathsf{T}$, and the matrices are
 \begin{align}
\bm{A}^{(x)}=
\begin{bmatrix}
 0 & 1 & 0 & 0 & 0 & 0 \\
 1 & 0 & 0 & 1 & 0 & 0 \\
 0 & 0 & 0 & 0 & 1 & 0 \\
 0 & \frac{2}{3} & 0 & 0 & 0 & 0 \\
 0 & 0 & \frac{1}{2} & 0 & 0 & 0 \\
 0 & -\frac{1}{3} & 0 & 0 & 0 & 0 \\
\end{bmatrix}
, \quad \bm{A}^{(y)}=
\begin{bmatrix}
 0 & 0 & 1 & 0 & 0 & 0 \\
 0 & 0 & 0 & 0 & 1 & 0 \\
 1 & 0 & 0 & 0 & 0 & 1 \\
 0 & 0 & -\frac{1}{3} & 0 & 0 & 0 \\
 0 & \frac{1}{2} & 0 & 0 & 0 & 0 \\
 0 & 0 & \frac{2}{3} & 0 & 0 & 0 \\
\end{bmatrix},
 \end{align}
 and $\bm{P}=\text{diag}(0, 0, 0, 1, 1, 1)$.
 On taking the Fourier transformation of the rewritten system, we obtain the matrix
\begin{align}
\bm{A}(\bm{k})=
\begin{bmatrix}
 0 & \mathbbm{i} k_x & \mathbbm{i} k_y & 0 & 0 & 0 \\
 \mathbbm{i} k_x & 0 & 0 & \mathbbm{i} k_x & \mathbbm{i} k_y & 0 \\
 \mathbbm{i} k_y & 0 & 0 & 0 & \mathbbm{i} k_x & \mathbbm{i} k_y \\
 0 & \frac{2 \mathbbm{i} k_x}{3} & -\frac{1}{3} (\mathbbm{i} k_y) & 1 & 0 & 0 \\
 0 & \frac{\mathbbm{i} k_y}{2} & \frac{\mathbbm{i} k_x}{2} & 0 & 1 & 0 \\
 0 & -\frac{1}{3} (\mathbbm{i} k_x) & \frac{2 \mathbbm{i} k_y}{3} & 0 & 0 & 1 \\
\end{bmatrix},
\end{align}
for which the symbol turns out to be
 \begin{align}
 s(\bm{k})=\frac{1}{2}(k_x^2+k_y^2)^2=\frac{1}{2} k^4,
 \end{align}
 where $k=\sqrt{k_x^2+k_y^2}$.
 In order to find the fundamental solution $\phi$ associated with the above symbol (such that $s(\bm{\nabla})[\phi]=\delta$), we utilize the definition~\eqref{ft} and~\eqref{ftinv} for the Biharmonic equation $\Delta^2 \phi =\delta$
whose fundamental solution in two dimensions is given by~\cite{CAO1994}
\begin{align}
\phi=\frac{r^2\,(\ln{r}-1)}{8\pi},
\end{align}
where $r=\sqrt{x^2+y^2}$.
This fundamental solution $\phi$ corresponds to the fundamental solution associated with the symbol for Stokes' equations and $\mathcal{F}^{-1}(1/k^4)=\phi$.
Applying the Fourier transformation [defined by Eq.~\eqref{ft}] to the Biharmonic equation  $\Delta^2 \phi =\delta$, we obtain
\begin{align}
 (-k_x^2-k_y^2)^2 \hat{\phi}=k^4\hat{\phi}=1 \quad\implies\quad
\hat{\phi}=\frac{1}{k^4}.
\end{align}
Taking inverse Fourier transformation,
\begin{align}
\mathcal{F}^{-1}\left(\frac{1}{k^4}\right)=\phi=\frac{r^2\, (\ln{r}-1)}{8\pi}.
\end{align}
Utilizing the above inverse Fourier transformation and the fundamental solution $\phi$, we thus obtain the complete fundamental solution for $\bm{U}$
\begin{align}
\hat{\bm{U}}=\frac{\bm{\mathcal{A}}(\bm{k})}{s(\bm{k})}\bm{S}=\frac{2}{k^4} \bm{\mathcal{A}}(\bm{k}) \bm{S} \implies \bm{U}=2 \bm{\mathcal{A}}(\bm{\nabla})[\phi] \bm{S},
\end{align}
where the adjugate matrix in operator form reads
\begin{align}\label{stokes_op_matrix}
\bm{\mathcal{A}}(\bm{\nabla})=
\scalebox{0.8}{\(
\begin{bmatrix}
 \frac{\Delta ^2}{3} & \frac{\Delta  \partial _x}{2} & \frac{\Delta  \partial _y}{2} & -\frac{1}{2}  \partial _x^2 \Delta &  -\partial _x \partial _y \Delta  & -\frac{1}{2} \partial _y^2 \Delta  \\[2mm]
 \frac{\Delta  \partial _x}{2} & -\partial _y^2 & \partial _x \partial _y & \partial _x \partial _y^2 & \partial _y^3-\partial _x^2 \partial _y & -\partial _x \partial _y^2 \\[2mm]
 \frac{\Delta  \partial _y}{2} & \partial _x \partial _y & -\partial _x^2 & -\partial _x^2 \partial _y & \partial _x (\partial _x^2-\partial _y^2)  & \partial _x^2 \partial _y \\[2mm]
 -\frac{1}{6} \Delta  \left(2 \partial _x^2-\partial _y^2\right) & \partial _x \partial _y^2 & -\partial _x^2 \partial _y & \frac{1}{2} \left(\partial _x^4+\partial _y^4\right) & \partial _x \partial _y (\partial _x^2-\partial _y^2) & \partial _x^2 \partial _y^2 \\[2mm]
 -\frac{1}{2} \Delta  \partial _x \partial _y & -\frac{1}{2} \partial _y (\partial _x^2-\partial _y^2) & \frac{1}{2} \partial _x (\partial _x^2-\partial _y^2) & \frac{1}{2} \partial _x \partial _y (\partial _x^2-\partial _y^2) & 2 \partial _x^2 \partial _y^2 & -\frac{1}{2} \partial _x \partial _y (\partial _x^2-\partial _y^2) \\[2mm]
 \frac{1}{6} \Delta  \left(\partial _x^2-2 \partial _y^2\right) & -\partial _x \partial _y^2 & \partial _x^2 \partial _y & \partial _x^2 \partial _y^2 & \partial _x \partial _y^3-\partial _x^3 \partial _y & \frac{1}{2} \left(\partial _x^4+\partial _y^4\right) \\
\end{bmatrix}
\)},
\end{align}
where $\Delta \equiv \partial_x^2+\partial_y^2$ represents the Laplacian operator. Applying the adjugate matrix~\eqref{stokes_op_matrix} to the fundamental solution $\phi$, we obtain the matrix containing basis functions used to approximate the overall solution via superposition, i.e.
\begin{align}
\label{fsol_stokes}
\bm{\mathfrak{A}}_{\text{Stokes}}(\bm{r})=
\scalebox{0.8}{\(
\begin{bmatrix}
 0 & \frac{x}{2 \pi  r^2} & \frac{y}{2 \pi  r^2} & \frac{x^2-y^2}{2 \pi r^4} & \frac{2 x y}{\pi r^4} & \frac{y^2-x^2}{2 \pi r^4} \\[2mm]
 \frac{x}{2 \pi  r^2} & -\frac{r^2 \log (r^2)-r^2}{4 \pi r^2} & \frac{x y}{2 \pi  r^2} & \frac{x (x^2-y^2)}{2 \pi r^4} & \frac{2 x^2 y}{\pi r^4} & -\frac{x (x^2-y^2)}{2 \pi r^4} \\[2mm]
 \frac{y}{2 \pi  r^2} & \frac{x y}{2 \pi  r^2} & -\frac{r^2 \log (r^2)+x^2-y^2}{4 \pi  r^2} & \frac{y (x^2-y^2)}{2 \pi r^4} & \frac{2 x y^2}{\pi r^4} & \frac{y (y^2-x^2)}{2 \pi r^4} \\[2mm]
 \frac{x^2-y^2}{2 \pi r^4} & \frac{x (x^2-y^2)}{2 \pi r^4} & \frac{y (x^2-y^2)}{2 \pi r^4} & \frac{x^4-6 x^2 y^2+y^4}{2 \pi  r^6} & \frac{4 x y (x^2-y^2)}{\pi  r^6} & -\frac{x^4-6 x^2 y^2+y^4}{2 \pi  r^6} \\[2mm]
 \frac{x y}{\pi  r^4} & \frac{x^2 y}{\pi  r^4} & \frac{x y^2}{\pi  r^4} & \frac{2 x y (x^2-y^2)}{\pi  r^6} & -\frac{x^4-6 x^2 y^2+y^4}{\pi  r^6} & -\frac{2 x y (x^2-y^2)}{\pi  r^6} \\[2mm]
 \frac{y^2-x^2}{2 \pi  r^4} & -\frac{x (x^2-y^2)}{2 \pi  r^4} & \frac{y (y^2-x^2)}{2 \pi  r^4} & -\frac{x^4-6 x^2 y^2+y^4}{2 \pi  r^6} & -\frac{4 x y (x^2-y^2)}{\pi  r^6} & \frac{x^4-6 x^2 y^2+y^4}{2 \pi  r^6} \\
\end{bmatrix}
\)},
\end{align}
where $\bm{\mathfrak{A}}_{\text{Stokes}}(\bm{r})\equiv 2\bm{\mathcal{A}}(\bm{\nabla})[\phi]$.
Now it remains to decide the entries of the vector $S$ which decides the Dirac-delta sourcing terms.
This choice will be discussed with an example setup in the following subsection.

\subsection{An Example Setup}
\begin{figure}[!ht]
\centering
\includegraphics[scale=0.65]{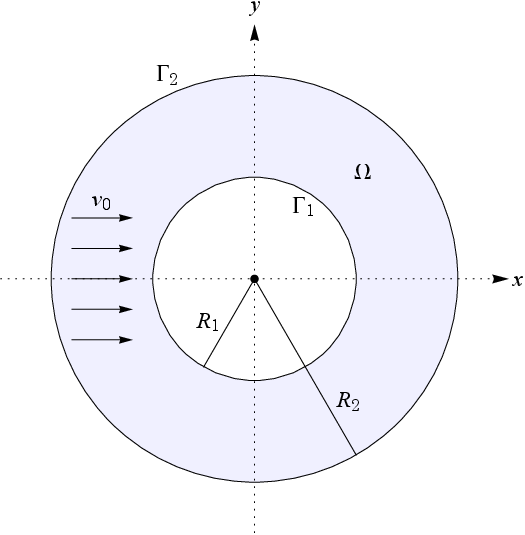}\quad
\includegraphics[scale=0.63]{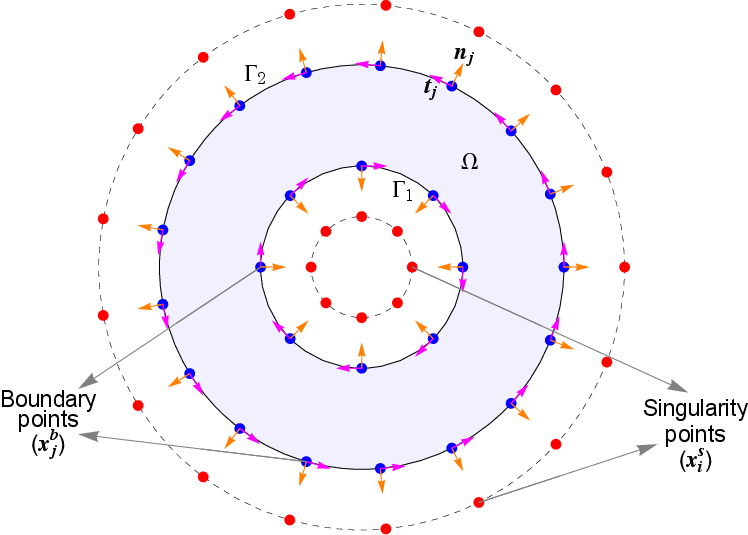}
 \caption{%
 \label{fig:stokes_schem} Stokes' flow between two cylinders (left) and the placement of boundary nodes and singularities in the MFS (right).}
\end{figure}

Let us consider a rarefied monatomic gas confined in between two infinitely long coaxial circular cylinders having radii $R_1$ and $R_2$ with $R_2>R_1$.
Owing to the axial symmetry, the problem can be investigated in two dimensions.
 A cross-sectional view of problem is depicted in the left panel of Fig.~\ref{fig:stokes_schem},
 where the flow domain is given by
\begin{align}
\Omega = \{ (x, y) \in \mathbb{R}^2 \mid R_1^2 \leq x^2 + y^2 \leq R_2^2 \},
\end{align}
with $\Gamma_1 = \{ (x, y)\in \mathbb{R}^2 \mid x^2 + y^2 = R_1^2 \}$ and $\Gamma_2 = \{ (x, y)\in \mathbb{R}^2 \mid x^2 + y^2 = R_2^2 \}$ denoting the inner and outer boundaries, respectively.
The inner cylinder is assumed to be impermeable with standard slip condition given by
\begin{align}
\label{bc_stokes_inner}
\bm{v}\cdot\bm{n}\big|_{\Gamma_1}=0 \quad \text{and} \quad
\bm{n} \cdot \bm{\sigma}\cdot\bm{t}\big|_{\Gamma_1}=-\zeta \bm{v}\cdot\bm{t}\big|_{\Gamma_1},
\end{align}
where $\bm{n}=(n_x,n_y)$ and $\bm{t}=(t_x,t_y)$ are the unit normal and tangent vectors on the inner boundary $\Gamma_1$ and $\zeta \in \mathbb{R}$ is the velocity-slip coefficient.
The outer cylinder enforces in- and out-flow boundary conditions, leading to
\begin{align}
\label{bc_stokes_outer}
\bm{v}\cdot\bm{n}\big|_{\Gamma_2}=v_0 \, n_x\big|_{\Gamma_2} \quad \text{and} \quad \bm{v}\cdot\bm{t}\big|_{\Gamma_2}=-v_0 \, n_y\big|_{\Gamma_2},
\end{align}
where $v_0 \in \mathbb{R}$ is the horizontal velocity. The boundary condition matrix constructed using~\eqref{bc_stokes_inner} and~\eqref{bc_stokes_outer}  for the unknown solution vector $\bm{U}=\big[
 p \;\; v_x \;\; v_y \;\; \sigma_{xx} \;\; \sigma_{xy} \;\; \sigma_{yy}
 \big]^\mathsf{T}$ is given by
\begin{align}
\bm{B}(\bm{x}^{\text{b}}) =
\begin{cases}
\begin{bmatrix}
 0 & n_x & n_y & 0 & 0 & 0 \\
 0 & \zeta  t_x & \zeta  t_y & n_x t_x & n_x t_y+n_y t_x & n_y t_y
\end{bmatrix}, & \text{if } \bm{x}^{\text{b}} \in \Gamma_1, \\[10pt]
\begin{bmatrix}
 0 & n_x & n_y & 0 & 0 & 0 \\
 0 &  t_x & t_y & 0  & 0 & 0
\end{bmatrix}, & \text{if } \bm{x}^{\text{b}} \in \Gamma_2.
\end{cases}
\end{align}
The right-hand-side vector is given by
 \begin{align}
\bm{g}(\bm{x}^{\text{b}}) =
\begin{cases}
\begin{bmatrix}
 0  \\
 0
\end{bmatrix}, & \text{if } \bm{x}^{\text{b}} \in \Gamma_1, \\[10pt]
\begin{bmatrix}
 v_0 \, n_x\\
 -v_0 \, n_y
\end{bmatrix}, & \text{if } \bm{x}^{\text{b}} \in \Gamma_2,
\end{cases}
\end{align}
where $\bm{x}^{\text{b}}$ represents the position of a point on the boundary of the cylinders.
In order to implement the MFS for the current setup, a total of $n_{\text{b}}$ boundary nodes are chosen on the boundaries $\Gamma_1$ and $\Gamma_2$.
Two concentric circular fictitious boundaries $\tilde{\Gamma}_1$ (inside $\Gamma_1$) and $\tilde{\Gamma}_2$ (outside $\Gamma_2$) are considered on which $n_{\text{s}}$ singularity points are placed as shown in the right panel of Fig.~\ref{fig:stokes_schem}.
The overall solution obtained from the MFS is then given by
\begin{align}
    \bm{U} (\bm{x})=\sum_{i=1}^{n_{\text{s}}} \bm{\mathfrak{A}}_{\text{Stokes}}(\bm{x}-\bm{x}_i^{\text{s}}) \bm{S}_i.
\end{align}
To find the unknown source strengths in $\bm{S}_i$, we split $\bm{S}=\bm M\bm{\mu}$, where $\bm{M}$ is a fixed matrix and $\bm{\mu}$ contains the deciding source strengths parameters.


\subsection{\label{sec:stokes_choiceM}Choice of the matrix \texorpdfstring{$\bm{M}$}{M}}
The main task in the MFS is to calculate the unknown source strengths using the boundary conditions.
For the classical Stokeslet approach, where a point force vector is included in the momentum balance equation, the corresponding matrix $\bm{M}$ is given by
 \begin{align}
 \label{M_stokeslet}
 \bm M=
\begin{bmatrix}
 0 & 1 & 0 & 0 & 0 & 0 \\
 0 & 0 & 1 & 0 & 0 & 0 \\
\end{bmatrix}
^\mathsf{T} ,
 \end{align}
and $\bm{\mu}=\big[
    \mu_1 \;\; \mu_2
\big]^\mathsf{T}$ represents the point force associated with the singularity.
Alternatively, one may introduce source terms into any of the Eqs.~\eqref{stokes1}--\eqref{stokes3}, for instance, setting $\bm{M}=I_6$, where $I_6$ is the $6 \times 6$ identity matrix, corresponds to Dirac delta source terms in all governing equations.
Nevertheless, while working with large and complex system of linear partial differential equation, it is not trivial to choose the non-zero entries in the vector $\bm{S}$ as the choice significantly affects the results.
We propose the choice of the matrix $\bm{M}$ to be dependent of the boundary conditions by fixing $\bm{M}(\bm{x}^{\text{b}})=\bm{B}(\bm{x}^{\text{b}})^\mathsf{T}$, which yields the boundary condition
\begin{align}
\label{bcMFS_stokes}
\bm{B}(\bm{x}^{\text{b}}) {\bm{\mathfrak{A}}_{\text{Stokes}}} \bm{B}(\bm{x}^{\text{b}})^\mathsf{T} \bm{\mu}=\bm g(\bm{x}^{\text{b}}),
\end{align}
 for any boundary point $\bm{x}^{\text{b}}$.
This choice of $\bm{M}$ is advantageous as it gives a symmetric structure to the overall system and keeps the number of source parameters in $\bm{\mu}$ equal to the number of boundary conditions at each node and yields a square system when the number of boundary nodes and source points are the same ($n_{\text{b}}=n_{\text{s}}$).
The system~\eqref{bcMFS_stokes} is evaluated at each boundary node for determining the source parameters in $\bm{\mu}$ corresponding to each singularity point.
This results in a large linear system
\begin{align}
    \bm{B}(\bm{x}_j^{\text{b}})\sum_{i=1}^{n_{\text{s}}} \bm{\mathfrak{A}}_{\text{Stokes}}(\bm{r}_{ij}) \bm{B}(\bm{x}_i^{\text{b}})^\mathsf{T} \bm{\mu}_i =\bm{g}(\bm{x}_j^{\text{b}}) , \quad j=1,2,\dots,n_{\text{b}}(=n_{\text{s}}),
\end{align}
where $\bm{r}_{ij}=\bm{x}_j^{\text{b}}-\bm{x}_i^{\text{s}}$ is the relative distance and $\bm{\mu}_i$ denotes the vector containing unknown source parameters corresponding to $i$th singularity point.
The complete linear system can be denoted by $\bm{\mathcal{L}}\bm{\Lambda}=\bm{\mathcal{G}}$, where $\bm{\mathcal{L}}$ is the $2n_{\text{b}} \times2 n_{\text{s}}$ collocation matrix and the unknown source strength vector is $\bm{\Lambda}=\big[
    \mu_{1_1} \;\; \mu_{1_2} \;\; \mu_{2_1} \;\; \mu_{2_2} \;\; \dots \;\; \mu_{{n_{\text{s}}}_1} \;\; \mu_{{n_{\text{s}}}_2}
\big]^\mathsf{T}$.
After calculating the unknown parameters in $\bm{\mu}$, one can approximate any flow variable by using the superposition $\bm U (\bm{x}) =\sum_{i=1}^{n_{\text{s}}}\bm{\mathfrak{A}}_{\text{Stokes}}(\bm{r}_i) \bm B(\bm{x}_i)^\mathsf{T} \bm{\mu}^i$, where $\bm{r}_i=\bm{x}-\bm{x}_i^{\text{s}}$ for any vector $\bm{x}$ in the computational flow domain.
For instance, the $x$-component of velocity $v_x$ can be calculated---using the second row of $\bm{\mathfrak{A}}_{\text{Stokes}}$ given in~\eqref{fsol_stokes}---as

\begin{align}\label{vx_stokes}
v_x=\sum_{i=1}^{n_{\text{s}}}
\begin{bmatrix}
 \frac{x_i}{2 \pi  r_i^2} & -\frac{r_i^2 (2 \log r_i-1)}{4 \pi r_i^2} & \frac{x_i y_i}{2 \pi  r_i^2} & \frac{x_i (x_i^2-y_i^2)}{2 \pi r_i^4} & \frac{2 x_i^2 y_i}{\pi r_i^4} & -\frac{x_i (x_i^2-y_i^2)}{2 \pi r_i^4} \\
\end{bmatrix}
\bm{B}(\bm{x}_i^{\text{b}})^\mathsf{T}
\begin{bmatrix}
     \mu_1^i \\
     \mu_2^i
\end{bmatrix}.
\end{align}
\section{\label{sec:r13_mfs}Generic MFS for R13 equations}

Expressing the R13 equations~\eqref{mass_bal}--\eqref{m_bal} in the form~\eqref{system}, the unknown vector is \\
$\bm{U}= \left[\begin{smallmatrix}
p & v_x & v_y & \sigma_{xx} & \sigma_{xy} & \sigma_{yy} & \theta & q_x & q_y & m_{xxx} & m_{xxy} & m_{yyx} &  m_{yyy} & R_{xx} & R_{xy} & R_{yy}
\end{smallmatrix}\right]^\mathsf{T}$.
Applying Fourier transformation on the resulting system~\eqref{system} as done in Sec.~\ref{sec:fund_stokes}, the symbol for R13 system turns out to be
\begin{align}
s(\bm{k})=\gamma (k^2)^3 (k^2+\lambda_1)(k^2+\lambda_2)(k^2+\lambda_3),
\end{align}
where
\begin{align}
\gamma=\frac{3087 \mathrm{Kn}^8}{160},\quad
\lambda_1=\frac{3}{2\mathrm{Kn}^2},\quad \lambda_2=\frac{5}{9\mathrm{Kn}^2} ,\quad \lambda_3=\frac{5}{6\mathrm{Kn}^2}.
\end{align}
These three constants $\lambda_1,\lambda_2$ and $\lambda_3$ represent the three Knudsen layers\footnote{Knudsen layers are thin boundary regions in rarefied gas flows where non-equilibrium effects dominate due to gas-surface interactions. The constants $\sqrt{5/6}$, $\sqrt{3/2}$, and $\sqrt{5}/3$ correspond to eigenvalues governing exponential decay rates of Knudsen layer modes in the R13 equations. Three eigenvalues align with the prediction of three Knudsen layers by the R13 model~\cite{Struchtrup2005,RTS2013,TorrilhonARFM}}.
This symbol in the operator form reads
\begin{align}
s(\Delta)=\gamma (\Delta)^3 (\Delta-\lambda_1)(\Delta-\lambda_2)(\Delta-\lambda_3),
\end{align}
where $\Delta\equiv \partial_x^2+\partial_y^2$. Utilizing Eqs.\ \eqref{system_ft} and~\eqref{inverse_adjugate_symbol} gives an idea to compute the main fundamental solution $\Phi$ corresponding to the symbol.
In Fourier transformed coordinates,
\begin{align}
\label{hat_U_r13}
\hat{\bm{U}}=\frac{1}{\gamma (k^2)^3 (k^2+\lambda_1)(k^2+\lambda_2)(k^2+\lambda_3)} \bm{\mathcal{A}}(\bm{k}) \bm{S}.
\end{align}
It is convenient to get the inverse Fourier transform of $\hat{\bm{U}}$ if $1/s(\bm{k})$ is expressed in its partial fraction form:
\begin{align}
    \frac{1}{s(\bm{k})}=\frac{1}{\gamma} \left[\frac{\alpha_1}{(k^2)^3}+\frac{\alpha_2}{(k^2)^2}+\frac{\alpha_3}{k^2}+\frac{\alpha_4}{k^2+\lambda_1}+\frac{\alpha_5}{k^2+\lambda_2}+\frac{\alpha_6}{k^2+\lambda_3}\right].
\end{align}
The constants $\alpha_i$s can be computed in a straightforward way, and hence Eq.~\eqref{hat_U_r13} becomes
\begin{align}
\hat{\bm{U}}=\frac{1}{\gamma} \left[\frac{36 \,\mathrm{Kn}^6}{25 (k^2) ^3}-\frac{132 \mathrm{Kn} ^8}{25 (k^2) ^2}+\frac{8356 \,\mathrm{Kn} ^{10}}{625 k^2}-\frac{8 \, \mathrm{Kn} ^{10}}{17 \left(k^2 +\frac{3}{2 \mathrm{Kn} ^2}\right)}\right.\nonumber
\\
\left. +\frac{5832 \, \mathrm{Kn} ^{10}}{625 \left(k^2 +\frac{5}{6 \mathrm{Kn} ^2}\right)}-\frac{236196 \, \mathrm{Kn} ^{10}}{10625 \left(k^2 +\frac{5}{9 \mathrm{Kn} ^2}\right)}\right]\bm{\mathcal{A}}(\bm{k})\bm{S}.
\end{align}
In order to compute the complete fundamental solution  $\Phi$, it is easier to use the inverse Fourier transforms of the partial fraction terms using the preknown fundamental solutions of polyharmonic or Helmholtz operators~\cite{CAO1994}.
For any polyharmonic equation having the fundamental solution $\phi_n$ which satisfies $\Delta^n \phi_n=\delta$, its Fourier transformation is obtained by using the property $\mathcal{F}\left( \partial F/\partial x_i \right) =\mathbbm{i} k_i$, which yields
\begin{align}
\label{polyharmonic}
    (-1)^{n} k^{2n} \hat{\phi}_n =\hat{\delta} =1 \implies \hat{\phi}_n=\frac{(-1)^{n}}{k^{2n}}.
\end{align}
Analogously, for a Helmholtz equation having the fundamental solution $\psi_\lambda$ which satisfies $(\Delta-\lambda) \psi_\lambda=\delta$, the Fourier transformation yields
\begin{align}
\label{helmohltz}
    (-k^2-\lambda)\hat{\psi}_\lambda=\hat{\delta}=1 \implies \hat{\psi}_\lambda=-\frac{1}{k^2+\lambda}.
\end{align}
Utilizing Eq.~\eqref{polyharmonic} and the preknown fundamental solutions for polyharmonic operators~\cite{CAO1994}, one can obtain
\begin{align}
\mathcal{F}^{-1}\left(\frac{1}{k^2}\right)&=-\phi_1=-
\frac{\log r}{2\pi}, 
\\
\mathcal{F}^{-1}\left(\frac{1}{k^4}\right)&=\phi_2=
\frac{r^2(\log r-1)}{8\pi}, 
\\
\mathcal{F}^{-1}\left(\frac{1}{k^6}\right)&=-\phi_3=-
\frac{r^4(\log r-3/2)}{128\pi}. 
\end{align}
Using fundamental solution $\psi_\lambda$ for Helmholtz equation, and Eq.~\eqref{helmohltz}, we get
\begin{align}
\mathcal{F}^{-1}\left(\frac{1}{k^2+\lambda}\right)&=-\psi_\lambda=-
\frac{K_0(\sqrt{\lambda} r)}{2\pi}.
\end{align}
Here, $K_0$ denotes the modified Bessel function of the second kind of order zero.
Since inverse Fourier transformation is linear, the fundamental solution $\Phi$ is
\begin{align}
\Phi (r)=-\frac{4178\, \text{Kn}^{10} \log r}{625 \pi }-\frac{33\, \text{Kn}^8\, r^2(\log r-1)}{50 \pi }-\frac{9\, \text{Kn}^6\, r^4(\log r-3/2)}{800 \pi }
\nonumber \\
+\frac{2916\, \text{Kn}^{10} K_0\left(\sqrt{\frac{5}{6}} \frac{r}{\text{Kn}}\right)}{625 \pi }-\frac{4\, \text{Kn}^{10} K_0\left(\sqrt{\frac{3}{2}} \frac{r}{\text{Kn}} \right)}{17 \pi }-\frac{118098\, \text{Kn}^{10} K_0\left(\frac{\sqrt{5}}{3}  \frac{r}{\text{Kn}}\right)}{10625 \pi }.
\end{align}
Taking the inverse Fourier transform in Eq.~\eqref{hat_U_r13}, we obtain the fundamental solution for the R13 equations as
\begin{align}
\bm{U}(\bm{r})=\frac{1}{\gamma}\bm{\mathcal{A}}(\bm{\nabla})[\Phi] \bm{S}=\bm{\mathfrak{A}}_{\text{R13}} (\bm{r}) \bm M \bm{\mu}.
\end{align}

 The matrix $\bm{\mathfrak{A}}_{\text{R13}}$ incorporates all fundamental solutions that contribute to approximating the complete numerical solution of any given problem.
  In the R13 system, different choices of the matrix  $\bm{M}$ allow for varying degrees of freedom. The choice can be made independent of the specific problem by setting $\bm{M}(\bm{x^{\text{b}}})=\bm{B}(\bm{x^{\text{b}}})^\mathsf{T}$.
  Here $\bm{B}(\bm{x^{\text{b}}}) \in \mathbb{R}^{6\times 16}$ boundary conditions matrix is constructed using boundary conditions~\eqref{bc_vn}--\eqref{bc_mntt}.
  With this choice (as also discussed in Sec.~\ref{sec:stokes_choiceM}), the unknown source strengths corresponding to the $i$th singularity $\bm{\mu}_i\in \mathbb{R}^{6}$ is calculated by solving the linear system
  \begin{align}
    \bm{B}(\bm{x}_j^{\text{b}})\sum_{i=1}^{n_{\text{s}}} \bm{\mathfrak{A}}_{\text{R13}}(\bm{r}_{ij}) \bm{B}(\bm{x}_i^{\text{b}})^\mathsf{T} \bm{\mu}_i =\bm{g}(\bm{x}_j^{\text{b}}) , \quad j=1,2,\dots,n_{\text{b}}(=n_{\text{s}}).
\end{align}
This linear system can be expressed as $\bm{\mathcal{L}}\bm{\Lambda}=\bm{\mathcal{G}}$, where $\bm{\mathcal{L}}$ is the $6 n_{\text{b}}\times 6n_{\text{b}}$ collocation matrix (due to $6$ boundary conditions associated with each boundary node and $6$ source strengths associated with each singularity).
Further, $\bm{\Lambda}\in \mathbb{R}^{6n_b}$ is the unknown vector (containing source strengths $\bm{\mu}_i$) and $,\bm{\mathcal{G}} \in \mathbb{R}^{6n_b}$ is the right-hand-side vector containing boundary properties $\bm{g}(\bm{x}_j^{\text{b}})$.
The numerical solution approximated by the MFS at any point $\bm{x}$ in the domain is determined by
\begin{align}
    \bm{U} (\bm{x})=\sum_{i=1}^{n_{\text{s}}} \bm{\mathfrak{A}}_{\text{R13}}(\bm{x}-\bm{x}_i^{\text{s}}) \bm{B}(\bm{x}_i^{\text{b}})^\mathsf{T} \bm{\mu}_i.
\end{align}

\section{Results and Discussion\label{sec:analytic}}
To validate our code for the generic MFS for the R13 equations, we compare the results obtained from the MFS with an analytical solution for a rarefied gas flow confined between two coaxial cylinders. Additionally, we examine the influence of various parameters on the accuracy of the numerical method.
\subsection{Problem description}
\begin{figure}[!t]
\centering
\includegraphics[scale=0.7]{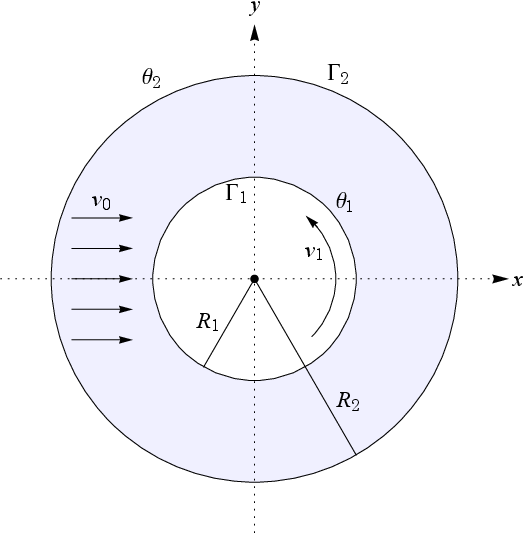} 
\caption{\label{fig:schematic2} Schematic of the cross-section of rarefied gas confined between two coaxial cylinders where the inner cylinder is rotating anticlockwise.}
\end{figure}

We consider a rarefied monatomic gas confined between two infinitely long coaxial circular cylinders. The dimensionless radii of the inner and outer cylinders are $R_1=1$ and $R_2=2$, respectively, with the inner and outer boundaries denoted by $\Gamma_1$ and $\Gamma_2$, respectively, as depicted in Fig~\ref{fig:schematic2}.
The outer cylinder serves as an inflow and outflow boundary, with normal component of velocity $\bm{v}^{\mathrm{w}}\cdot\bm{n} \big|_{\Gamma_2}=v_0 n_x\big|_{\Gamma_2}$ and tangential component $\bm{v}^{\mathrm{w}}\cdot\bm{t} \big|_{\Gamma_2}=-v_0 n_y\big|_{\Gamma_2}$ in the boundary conditions~\eqref{bc_vn}--\eqref{bc_mntt}.
To introduce additional complexity, the inner cylinder is assumed to be rotating with a tangential velocity, given by $\bm{v}^{\mathrm{w}}\cdot\bm{t} \big|_{\Gamma_1}=-v_1$.
The temperatures of the inner and outer cylinders are fixed at $\theta^{\mathrm{w}} \big|_{\Gamma_1}=\theta_1=1$ and $\theta^{\mathrm{w}} \big|_{\Gamma_2}=\theta_2=2$, respectively.
 The velocity prescription coefficient at inner cylinder is $\epsilon^{\mathrm{w}}\big|_{\Gamma_1}=10^{-5}$, while that on outer cylinder is $\epsilon^{\mathrm{w}}\big|_{\Gamma_2}=1$.
Furthermore, we fix $v_0=v_1=1$ and $p^{\mathrm{w}}\big|_{\Gamma_1}=p^{\mathrm{w}}\big|_{\Gamma_2}=0$.

\subsection{Validation with analytic solution}
\begin{figure}[!t]
\centering
\includegraphics[scale=0.7]{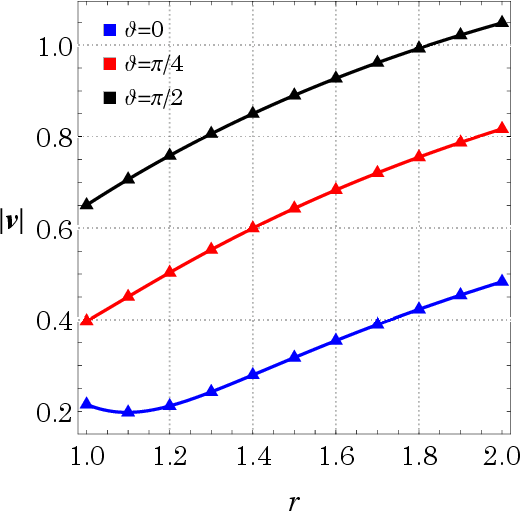} \quad
\includegraphics[scale=0.7]{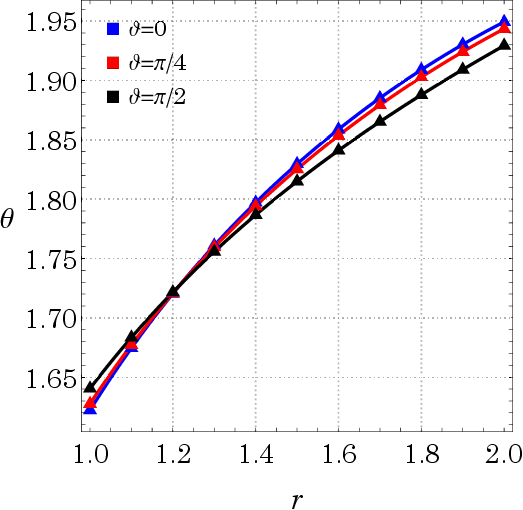}
\caption{%
Variation of the speed (left panel) and temperature (right panel) in the gap between the two cylinders. The solid blue, red and black lines denote the analytic results of the R13 model along $\vartheta=0,\pi/4$ and $\pi/2$, respectively. The corresponding blue, red and black (triangle) symbols denote the results obtained from the MFS for $\mathrm{Kn}=0.5$.}\label{fig:comp_analytic}
\end{figure}

The details for obtaining the analytic solution to this problem are provided in Appendix~\ref{app:A}.
To validate the code, we plot the speed of gas varying with radial gap between the two cylinders along different directions in the left panel of Fig.~\ref{fig:comp_analytic}.
The solid blue, red and black lines indicate the results obtained from the analytic solution of the R13 model for the azimuthal angles $\vartheta=0,\pi/4$ and $\pi/2$, respectively, whereas the symbols (triangles) represent the corresponding results obtained from the MFS for $\mathrm{Kn}=0.5$.
The right panel of Fig.~\ref{fig:comp_analytic} illustrates the variation in temperature with respect to the radial gap along different angles.
We observe an excellent agreement between the results obtained from the MFS and those from the analytic solution for both speed and temperature.
The complete source code for the generic MFS and the analytical solution for the R13 equations has been made publicly accessible\footnote{\href{https://github.com/himanshikhungar/R13_MFS}{https://github.com/himanshikhungar/R13\_MFS}}~\cite{himanshi_2025_15279486}.
For a better analysis, we measure the accuracy of the generic MFS in the following subsection using the standard relative error in the $L^2$ norm
\begin{align}
\epsilon_{L^2}=\frac{\Vert f_{\text{MFS}}-f_{\text{exact}}\Vert_{L^2(\Omega)}}{\Vert f_{\text{exact}}\Vert_{L^2(\Omega)}},
\end{align}
where $f_{\text{MFS}}$ denotes the numerical solution obtained with the MFS and $f_{\text{exact}}$ denotes the corresponding analytic solution.

%
%
%

\subsection{Choice of Parameters}
The accuracy of the MFS solution is highly dependent on key parameters, namely the numbers of boundary and source points, and the location of source points outside the computational domain.
To systematically analyze the error and justify the choice of these parameters, we define a grid spacing parameter $d$, which determines the distance between two consecutive boundary points.
 A smaller $d$ results in a higher number of boundary points and vice versa.
Given the grid spacing parameter $d$, the number of boundary points placed on the circumference of a circle of radius $R$ is computed as $n_{\text{b}} = \lfloor 2\pi R / d \rfloor$, where $\lfloor \cdot \rfloor$ denotes the floor function.
As previously mentioned, we set the number of boundary points equal to the number of source points to construct a square linear system using the relation $\bm{M}=\bm B(\bm{x})^\mathsf{T}$.

\begin{figure}[!t]
\centering
\includegraphics[scale=0.65]{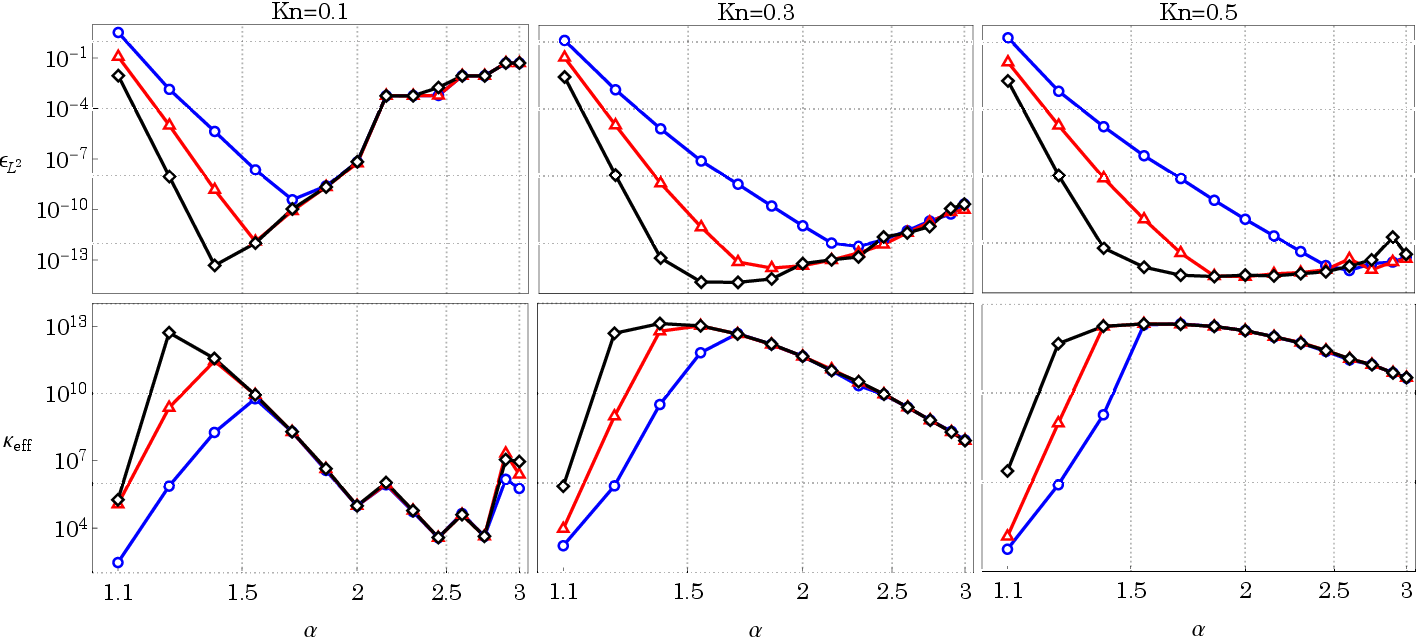}
 \includegraphics[scale=0.7]{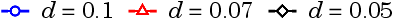}
\caption{Variation in $L^2$ error in velocity $\epsilon_{L^2}$ and effective condition number $\kappa_{\mathrm{eff}}$ with respect to the dilation parameter $\alpha$ for different values of grid spacing $d$ and $\bm{M}=\bm B(\bm{x})^\mathsf{T}$}\label{fig:err_keff}
\end{figure}

To determine an appropriate placement of source points, we introduce the dilation parameter $\alpha=R_1/R_{s_1}=R_{s_2}/R_2$ where $R_{s_1}$ and $R_{s_2}$ denote the radii of the inner and outer fictitious boundaries on which source points are placed.
A larger $\alpha$ corresponds to source points being positioned farther from the boundary and vice versa.
To evaluate the accuracy of the MFS, we compute the $L^2$ error in velocity $\epsilon_{L^2}$ for different values of $d$ and $\alpha$.
The top panels in Fig.~\ref{fig:err_keff} illustrate the variation in $\epsilon_{L^2}$ with respect to $\alpha$ for grid spacings $d \in \{0.1, 0.07, 0.05\}$ and Knudsen numbers $\mathrm{Kn}\in \{0.1,0.3,0.5\}$.
For a higher Knudsen number $\mathrm{Kn}=0.5$ (rightmost top panel), fewer boundary points ($d=0.1$) provide good accuracy when $\alpha$ is sufficiently large, meaning the source points are placed sufficiently far from the boundary.
In contrast, for $d=0.07$ and $d=0.05$, accurate results are achieved for $\alpha\gtrsim 1.7$ and $\alpha\gtrsim 1.5$, respectively.
This suggests that for computational efficiency, a smaller number of boundary points with more distant source points can be a viable choice.
However, for lower Knudsen numbers ($\mathrm{Kn}=0.1$ and $0.3$, leftmost and middle top panels), the accuracy depends more sensitively on the choice of boundary and source points.
The error is minimized only within a narrow range of $\alpha$, particularly for $\mathrm{Kn}=0.1$, indicating that source points should neither be too close nor too far from the boundary for an optimum accuracy.


The accuracy of the numerical solution depends strongly on the Knudsen number, which makes it challenging to determine where the source points should be placed, especially in the absence of an analytic solution.
One useful measure to guide this choice is the condition number that reflects how well the numerical system is posed. In the MFS, the corresponding linear system typically leads to a collocation matrix that is highly ill-conditioned.
To address this, several studies have examined the trade-off between numerical accuracy and matrix conditioning by evaluating a quantity known as the effective condition number~\cite{DML2009,CNYC2023,WL2011,HRG2023,HRG2025}.
The effective condition number provides a more reliable indicator of solution accuracy than the standard condition number, as it incorporates the right-hand-side vector.
For a linear system written as $\bm{\mathcal{L}}\bm{\Lambda} = \bm{\mathcal{G}}$, the singular value decomposition $\bm{\mathcal{L}} = \bm{U D V}^\mathsf{T}$ yields the smallest non-zero singular value $\sigma_m$.
This value is used to define the effective condition number as
\begin{align}
\kappa_{\text{eff}}=\frac{\Vert \bm{\mathcal{G}} \Vert_2}{\sigma_m \Vert \bm{\Lambda} \Vert_2}.
\end{align}
The bottom panel in Fig.~\ref{fig:err_keff} shows how the effective condition number $\kappa_{\text{eff}}$ varies with the dilation parameter $\alpha$ for three values of the grid spacing $d \in \{0.1, 0.07, 0.05\}$ and Knudsen numbers $\mathrm{Kn} \in \{0.1,0.3,0.5\}$.
A comparison with the corresponding top panel, which displays the error $\epsilon_{L^2}$, reveals an approximate inverse proportionality between $\kappa_{\text{eff}}$ and the error: in general, lower errors are associated with higher values of $\kappa_{\text{eff}}$.
However, the value of $\alpha$ corresponding to the minimum error does not always align precisely with the peak in $\kappa_{\text{eff}}$, indicating that the relationship is not strictly proportional.
This observation suggests that the source point placement should be chosen to strike a balance---achieving sufficiently high $\kappa_{\text{eff}}$ while also minimizing the numerical error. Based on this reasoning, we select $\alpha = 1.5$ and $d = 0.05$ for our computations.


\subsection{\label{sec:r13_choiceM}Choice of the matrix \texorpdfstring{$\bm{M}$}{M} }
\begin{figure}[!t]
\centering
\includegraphics[height=4cm]{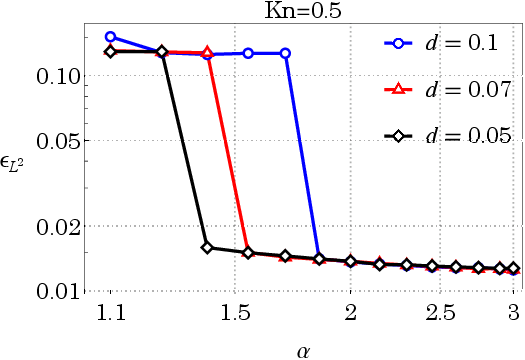} \quad
\includegraphics[height=4cm]{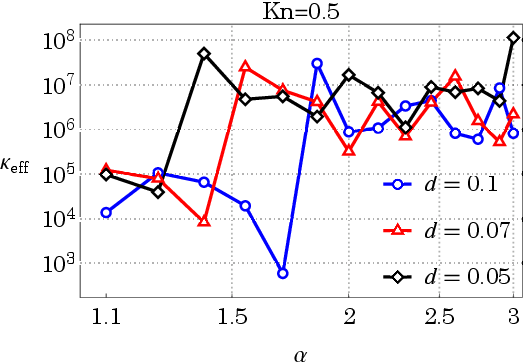}
\caption{\label{fig:6dof}Variation in $L^2$ error in velocity $\epsilon_{L^2}$ (left panel) and effective condition number $\kappa_{\mathrm{eff}}$ (right panel) with respect to dilation parameter $\alpha$ for $\bm{M}=\big[
    \bm{0}_{1\times 6} \;\; I_6 \;\; \bm{0}_{9\times 6}
\big]^\mathsf{T}$ for $\mathrm{Kn}=0.5$.}
\end{figure}

As discussed in Sec.~\ref{sec:intro}, previous studies on the MFS for rarefied gas flows formulated the fundamental solutions by imposing only a few degrees of freedom as Dirac-delta source terms in some governing equations and/or in some closure relations.
Ref.~\citep{CSRSL2017} derived the fundamental solutions for the R13 equations by including sourcing terms in the momentum, energy and stress balance equations.
 This choice ensured that the number of boundary conditions matched the number of unknown sources associated with each singularity.
 The fundamental solutions derived in~\citep{CSRSL2017} can also be deduced using the general matrix $\bm{\mathfrak{A}}_{\text{R13}}$ containing the fundamental solutions.
Setting $\bm{M}=\big[
    \bm{0}_{1\times 6} \;\; I_6 \;\; \bm{0}_{9\times 6}
\big]^\mathsf{T}$, leads to the parameter $\bm{\mu}= \big[
    \mu_1 \;\; \mu_2 \;\; \mu_3 \;\; \mu_4 \;\; \mu_5 \;\; \mu_6
\big]^\mathsf{T}$ with six degrees of freedom.
In this scenario, the linear system formed by implementing boundary conditions at each boundary node reads
 \begin{align}
    \bm{B}(\bm{x}_j^{\text{b}})\sum_{i=1}^{n_{\text{s}}} \bm{\mathfrak{A}}_{\text{R13}}(\bm{r}_{ij}) \bm{M} \bm{\mu}_i =\bm{g}(\bm{x}_j^{\text{b}}) , \quad j=1,2,\dots,n_{\text{b}}.
\end{align}
While this approach was effective for the specific problem considered in~\citep{CSRSL2017}, this particular choice may not always yield accurate results.
To illustrate this, Fig.~\ref{fig:6dof} shows the variation in the $L^2$ error in velocity (left panel) and the effective condition number $\kappa_{\text{eff}}$ (right panel) as functions of the dilation parameter $\alpha$ for $\mathrm{Kn}=0.5$.
As evident from the left panel of Fig.~\ref{fig:6dof}, the error remains large at all the locations of the source points, and the effective condition number does not exhibit any structured behavior.
Although not shown here, the errors remain high for all the considered Knudsen number values as well.
This suggests that the choice with six degrees of freedom does not perform well for the present problem.
A more suitable choice for the current study is to set $\bm{M}=\big[
    I_9 \;\; \bm{0}_{7\times 9}
\big]^\mathsf{T}$, which introduces nine degrees of freedom corresponding to mass, momentum, energy, stress, and heat balance equations.
In this case, the collocation matrix $\bm{\mathcal{L}}$ has dimensions $6n_{\text{b}}\times 9n_{\text{s}}$ and the corresponding linear system can be solved using the least squares method if $6n_{\text{b}}> 9n_{\text{s}}$ or one can fix $n_{\text{b}}$ and $n_{\text{s}}$ such that $6n_{\text{b}}= 9n_{\text{s}}$ and the resulting collocation matrix is square.
\begin{figure}[!t]
\centering
\includegraphics[height=4cm]{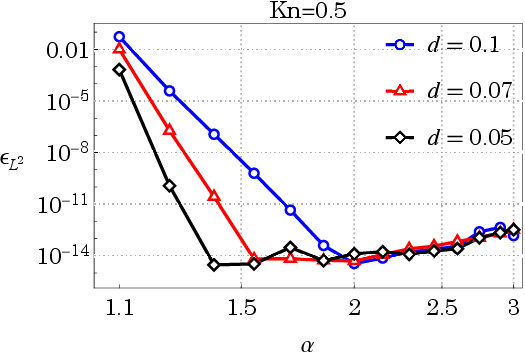} \quad
\includegraphics[height=4cm]{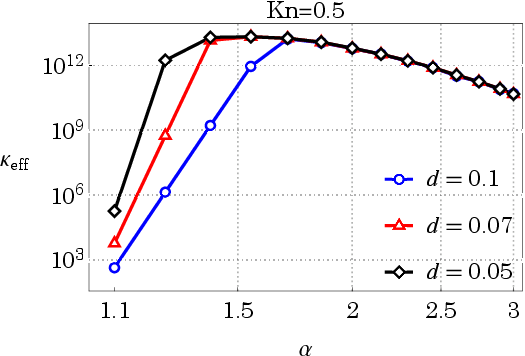}
\caption{\label{fig:9dof}Variation in $L^2$ error in velocity $\epsilon_{L^2}$ (left panel) and effective condition number $\kappa_{\mathrm{eff}}$ (right panel) with respect to dilation parameter $\alpha$ for $\bm{M}=\big[
    I_9 \;\; \bm{0}_{7\times 9}
\big]^\mathsf{T}$ for $\mathrm{Kn}=0.5$.}
\end{figure}
Figure~\ref{fig:9dof} illustrates the variation in $L^2$ error in velocity (left panel) and effective condition number (right panel) with the dilation parameter $\alpha$ for $\mathrm{Kn}=0.5$.
The behavior of both $\epsilon_{L^2}$ and $\kappa_{\text{eff}}$ closely resemble with those observed for $\bm{M}=\bm{B}(\bm{x})^\mathsf{T}$ in Fig.~\ref{fig:err_keff}.
Although not shown here, the resemblance exists for $\mathrm{Kn}=0.1$ and $0.3$ as well.
The comparison indicates that this choice of $\bm{M}=\big[
    I_9 \;\; \bm{0}_{7\times 9}
\big]^\mathsf{T}$ is more appropriate than $\bm{M}=\big[
    \bm{0}_{1\times 6} \;\; I_6 \;\; \bm{0}_{9\times 6}
\big]^\mathsf{T}$ for the present problem.
However, this choice of $\bm{M}$ cannot be guaranteed to work well for other problems.

\section{\label{sec:fem}Comparison with the FEM}
After validating the generic MFS framework for the R13 equations with an analytic solution, we now consider a problem for which an analytic solution is unknown.
The results are therefore compared to the results obtained from the FEM.
Furthermore, we observe the key differences and advantages of the MFS over FEM.
\subsection{Problem description}\label{sec:fem_problem_description}
In this scenario, a monatomic rarefied gas is considered to be confined between two noncoaxial infinitely long cylinders.
The circular cross-sections of the inner and outer cylinders have radii $R_1=1$ and $R_2=2$, respectively and centers at $(0,-0.25)$ and $(0,0)$, respectively. The boundaries are again denoted by $\Gamma_1$ and $\Gamma_2$, respectively.
The (dimensionless) temperatures on the inner and outer cylinders are fixed at $\theta^{\mathrm{w}} \big|_{\Gamma_1}=\theta_1=1$ and $\theta^{\mathrm{w}} \big|_{\Gamma_2}=\theta_2=2$, respectively.
Both the cylinders are assumed to be stationary ($\bm{v}^{\mathrm{w}}\big|_{\Gamma_1}=\bm{v}^{\mathrm{w}}\big|_{\Gamma_2}=0$) with the velocity prescription coefficient $\epsilon^{\mathrm{w}}\big|_{\Gamma_1}=\epsilon^{\mathrm{w}}\big|_{\Gamma_2}=0$ in the boundary conditions~\eqref{bc_vn}--\eqref{bc_mntt}. The flow is induced purely by the temperature difference.

\subsection{FEM for the R13 model}
In the FEM (see, e.g.,~\cite{doneaFiniteElementMethods2003} for an introduction focusing on flow equations), the equations are not solved pointwise for all \(\boldsymbol{x} \in \Omega\), but in an integral sense (weakly) on a triangulation \(\mathcal{T}_h\) of \(\Omega\) into \emph{finite elements} \(\tau \in \mathcal{T}_h\) (triangles in our case). Here, \(h \in \mathbb{R}\) denotes the maximum diameter of the elements. In contrast to the first-order system~\eqref{system}, we do not solve for all moments but restrict ourselves to the three balance laws~\eqref{mass_bal},~\eqref{mom_bal}, and~\eqref{energy_bal}, complemented by the two additional Eqs.~\eqref{sigma_bal} and~\eqref{q_bal}. The higher-order moments~\eqref{R_bal},~\eqref{m_bal}, and~\eqref{delta_bal} are directly inserted into these five equations, resulting in a field vector \(\bm{V}=\big[
p \;\; v_x \;\; v_y \;\; \sigma_{xx} \;\; \sigma_{xy} \;\; \sigma_{yy} \;\; \theta \;\; q_x \;\; q_y
\big]^\mathsf{T}\).
\begin{figure}[t]%
    \centering%
    \subfloat[%
      \(\mathcal{T}_1\) (\(h_{\max} \approx 0.932\)).\label{sfig:mesh1}%
    ]{%
    \includegraphics[height=0.245\linewidth]{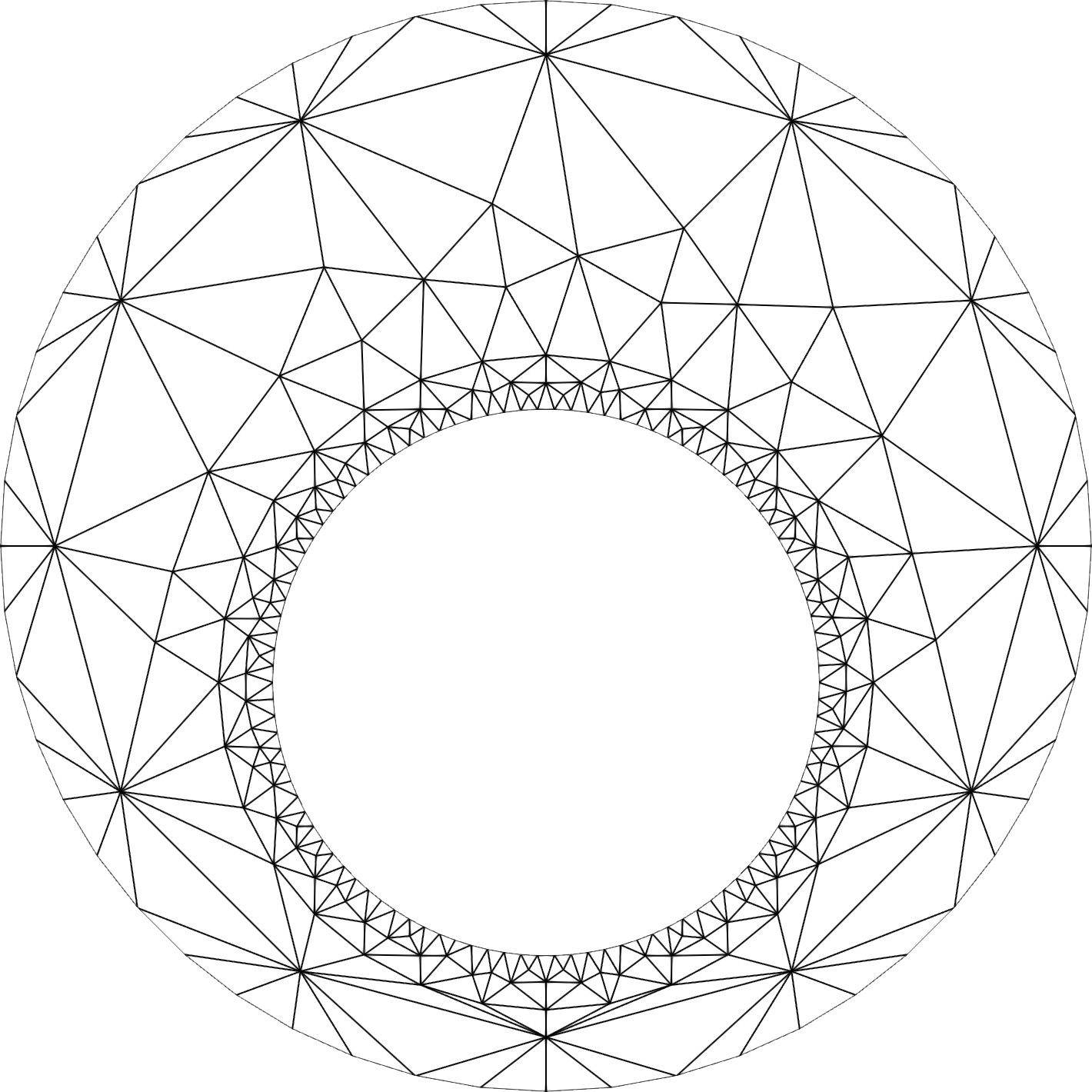}%
    }%
    \hfill%
    \subfloat[%
      \(\mathcal{T}_2\) (\(h_{\max} \approx 0.47\)).\label{sfig:mesh2}%
    ]{%
    \includegraphics[height=0.245\linewidth]{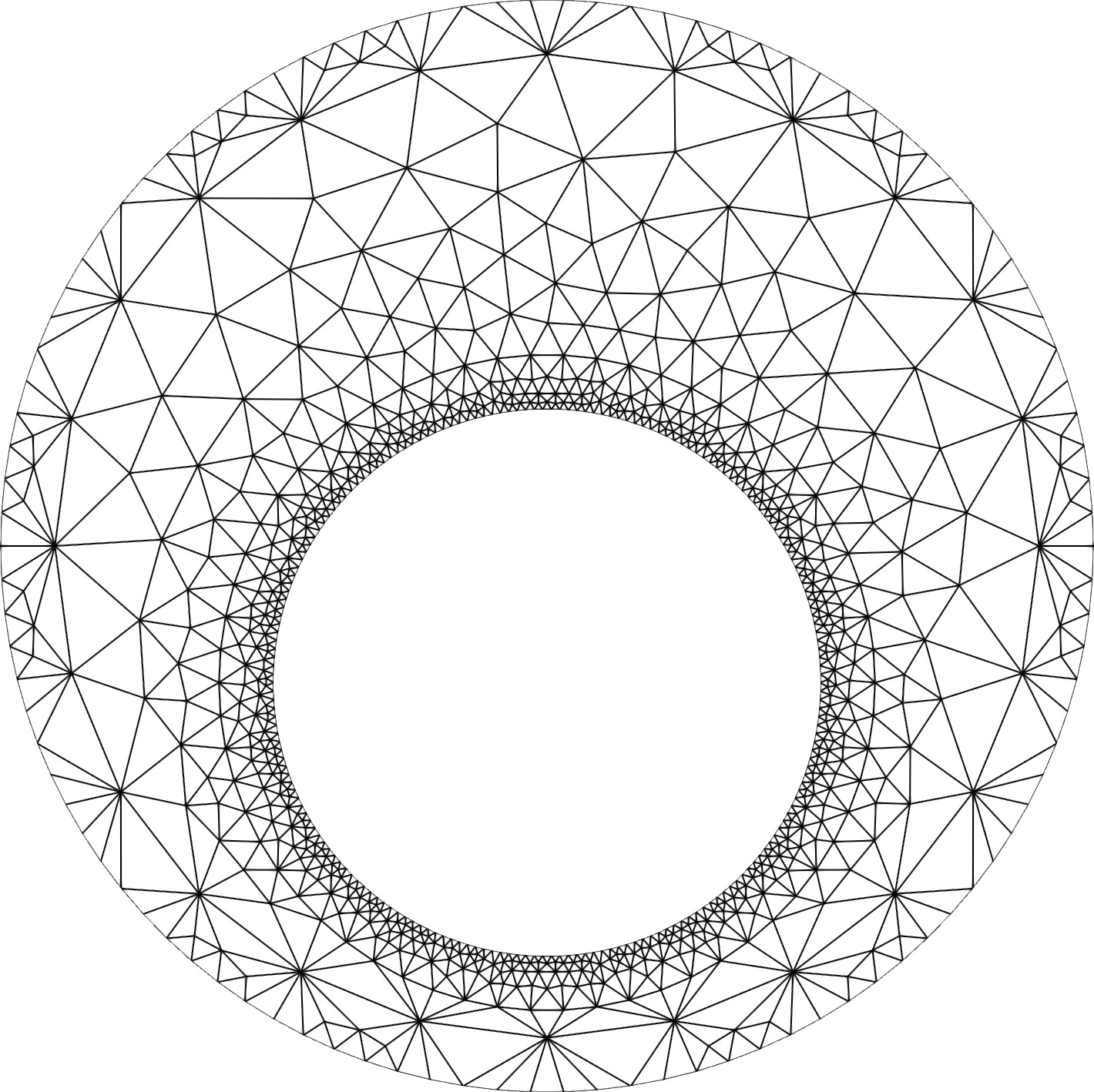}%
    }
    \hfill%
    \subfloat[%
      \(\mathcal{T}_3\) (\(h_{\max} \approx 0.281\)).\label{sfig:mesh3}%
    ]{%
    \includegraphics[height=0.245\linewidth]{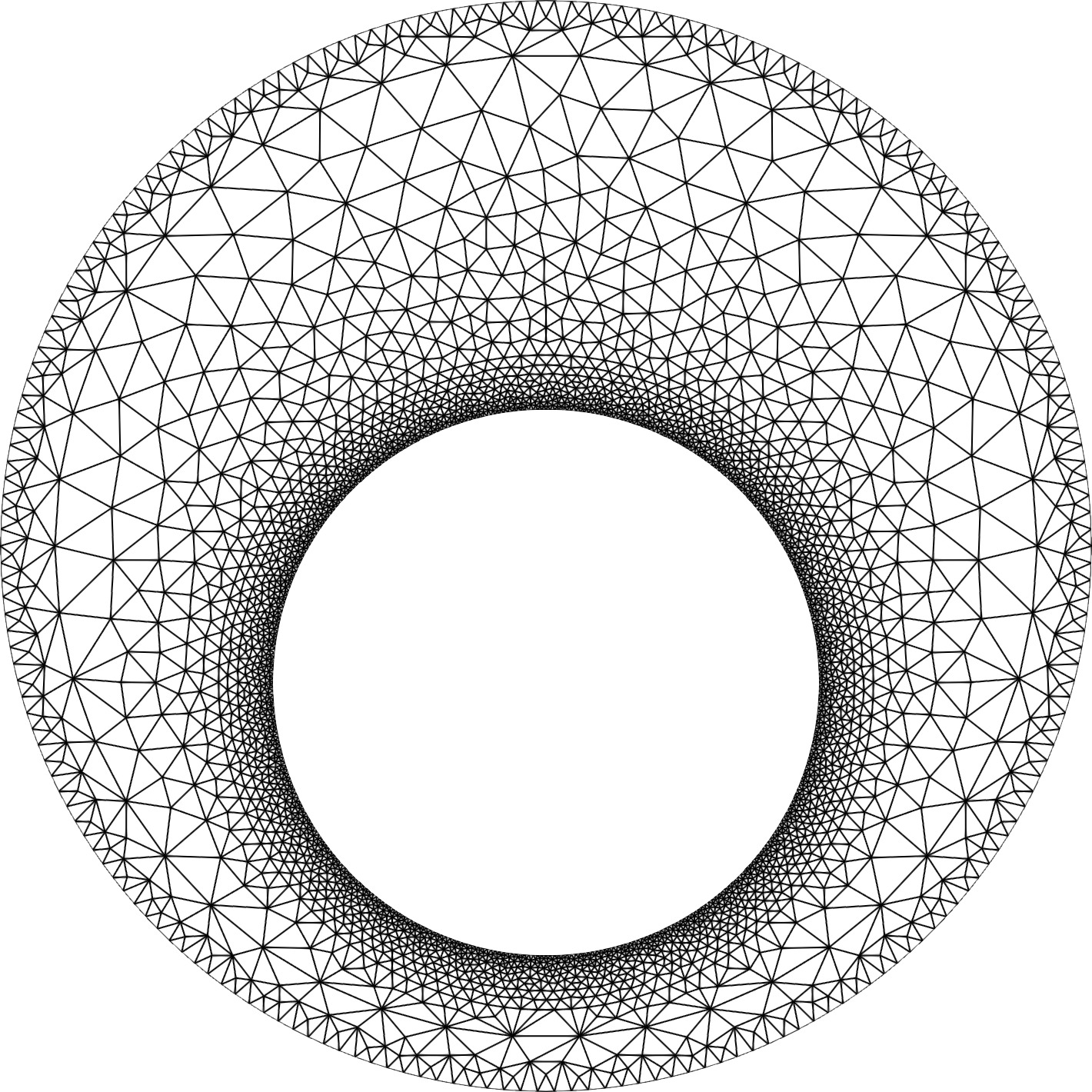}%
    }
    \hfill%
    \subfloat[%
      \(\mathcal{T}_4\) (\(h_{\max} \approx 0.16\)).\label{sfig:mesh4}%
    ]{%
    \includegraphics[height=0.245\linewidth]{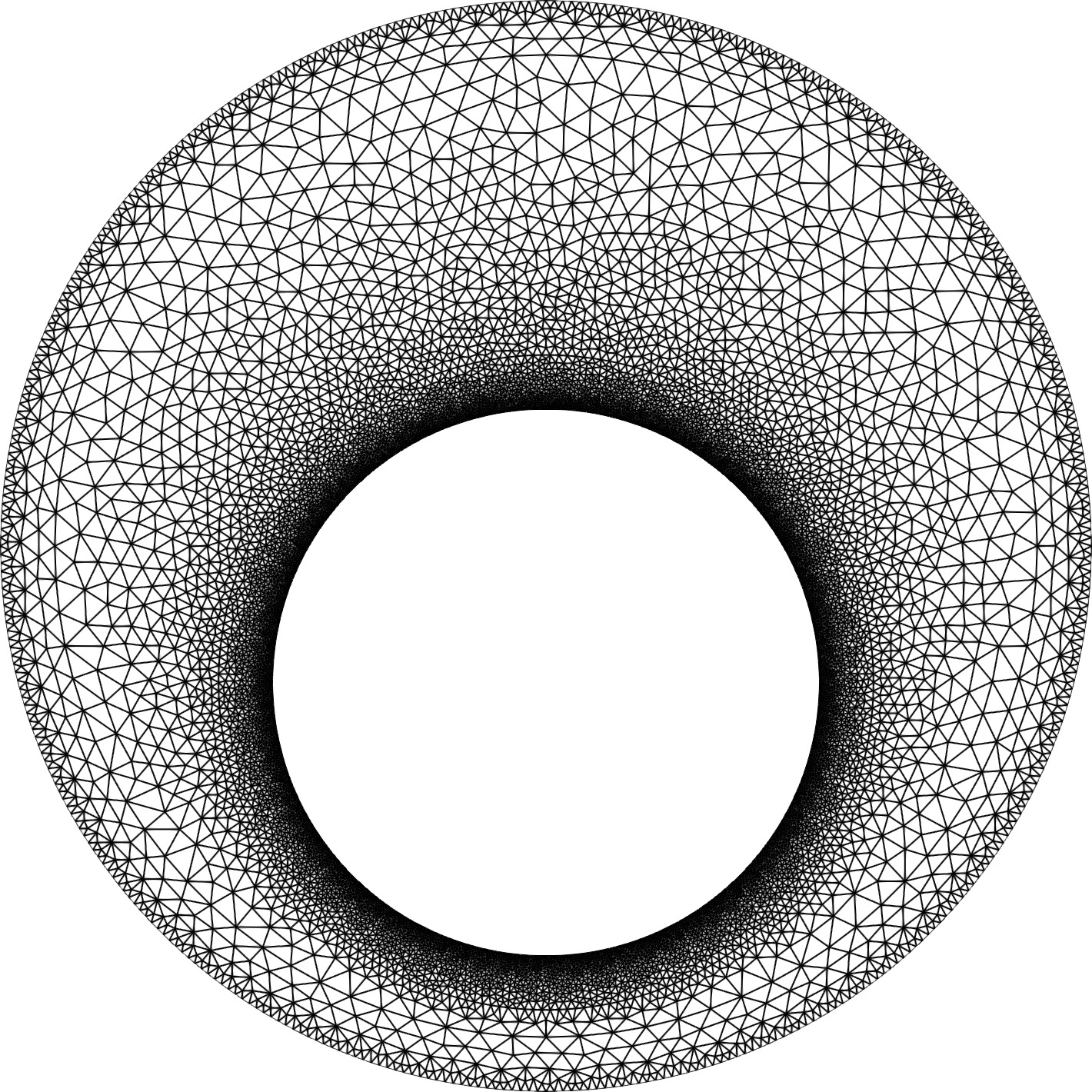}%
    }
    \caption{%
        Series of finite element meshes \(\mathcal{T}_i\) with decreasing mesh size \(h_{\max}\) for increasing \(i\).%
    }\label{fig:fem_meshes}%
\end{figure}%
\par
To obtain the weak formulation, we multiply each equation by corresponding test functions \(W = (\varphi_p, \dots)\), integrate over \(\Omega\), and apply integration by parts. This procedure lowers the order of differentiation and allows incorporating the boundary conditions~\eqref{bc_vn}--\eqref{bc_mntt}. In the Galerkin approach, the test functions are chosen from the same finite element space as the solution. An example is the weak formulation of the mass balance~\eqref{mass_bal}, where testing with \(\varphi_p : \Omega \to \mathbb{R}\) and integrating by parts yields
\begin{align}
  \int_\Omega \left( \bm{\nabla} \cdot \bm{v} \right) \varphi_p \, \mathrm{d}\boldsymbol{x}
  &=
  - \int_\Omega \bm{v} \cdot \bm{\nabla} \varphi_p \, \mathrm{d}\boldsymbol{x}
  + \int_\Gamma v_n \varphi_p \, \mathrm{d}l \nonumber \\
  &=
  - \int_\Omega \bm{v} \cdot \bm{\nabla} \varphi_p \, \mathrm{d}\boldsymbol{x}
  + \int_\Gamma \left( \epsilon^\mathrm{w} \tilde{\chi} \left( (p-p^\mathrm{w}) + \sigma_{nn} \right) + v_n^{\mathrm{w}} \right) \varphi_p \, \mathrm{d}l,
\end{align}
A reordering of terms for the unknowns and test functions leads to
\begin{equation}
  \int_\Omega \bm{v} \cdot \bm{\nabla} \varphi_p \, \mathrm{d}\boldsymbol{x}
  + \int_\Gamma \epsilon^\mathrm{w} \tilde{\chi} \left( p + \sigma_{nn}\right) \varphi_p \, \mathrm{d}l
  =
  -
  \int_\Gamma \left( v_n^{\mathrm{w}} - \epsilon^\mathrm{w} p^\mathrm{w} \right) \varphi_p \, \mathrm{d}l
  ,
  \label{eq_weakmass}
\end{equation}
which has to hold for all \(\varphi_p \in W\). Repeating these steps for all equations leads to a well-posed system~\cite{TT2021,lewintanWellPosednessR13Equations2025}. Finally, we discretize all functions in \(V\) and \(W\) by approximating them in the finite element space, i.e.\ as a linear combination of basis functions \(\phi_{\star,i}\) with coefficients \(c_i\), such that, for example,
\begin{equation}\label{eq-fem_ansatz}
    p(\boldsymbol{x}) = \sum_{i=1}^{N_p} c_{p,i} \phi_{p,i}(\boldsymbol{x}).
\end{equation}
We use stabilized first-order Lagrange elements, which are piecewise linear and globally continuous on the mesh. Inserting the ansatz~\eqref{eq-fem_ansatz} into the weak equations and evaluating the integrals via numerical quadrature, we obtain a linear system of equations:
\begin{equation}
    \boldsymbol{A}_{h} \boldsymbol{x}_{h} = \boldsymbol{b}_{h},
\end{equation}
where \(\boldsymbol{A}_{h} \in \mathbb{R}^{N \times N}\) is a sparse system matrix, \(\boldsymbol{x}_{h} \in \mathbb{R}^N\) contains the degrees of freedom of the solution (i.e.\ the vectors of coefficients \({\{c_{\star,i}\}}_{i=1}^{N_\star}\)), and \(\boldsymbol{b}_{h} \in \mathbb{R}^N\) is the right-hand side vector. The sparsity of \(\boldsymbol{A}_{h}\) results from the local support of the basis functions, i.e.\ \(\phi_{\star,i}\) is non-zero only on a small subset of elements \(\tau\).
\par
However, particularly for thermally induced flows as discussed in Sec.~\ref{sec:fem_problem_description}, a fine and locally refined mesh is required to accurately capture the characteristic flow features. For the test case, we generated a sequence of meshes \(\{\mathcal{T}_1,\dots,\mathcal{T}_7\}\) with decreasing maximal radii \(h_{\max}\). The first four of these meshes are shown schematically in Fig.~\ref{fig:fem_meshes} and illustrate the essential requirement of local refinement near the boundaries. For full reproducibility, the FEM source code along with all metadata is publicly available at~\cite{theisenFenicsR13TensorialMixed2025}.

\subsection{Results and Discussion}
\begin{figure}[!t]
    \centering
    \includegraphics[scale=0.65]{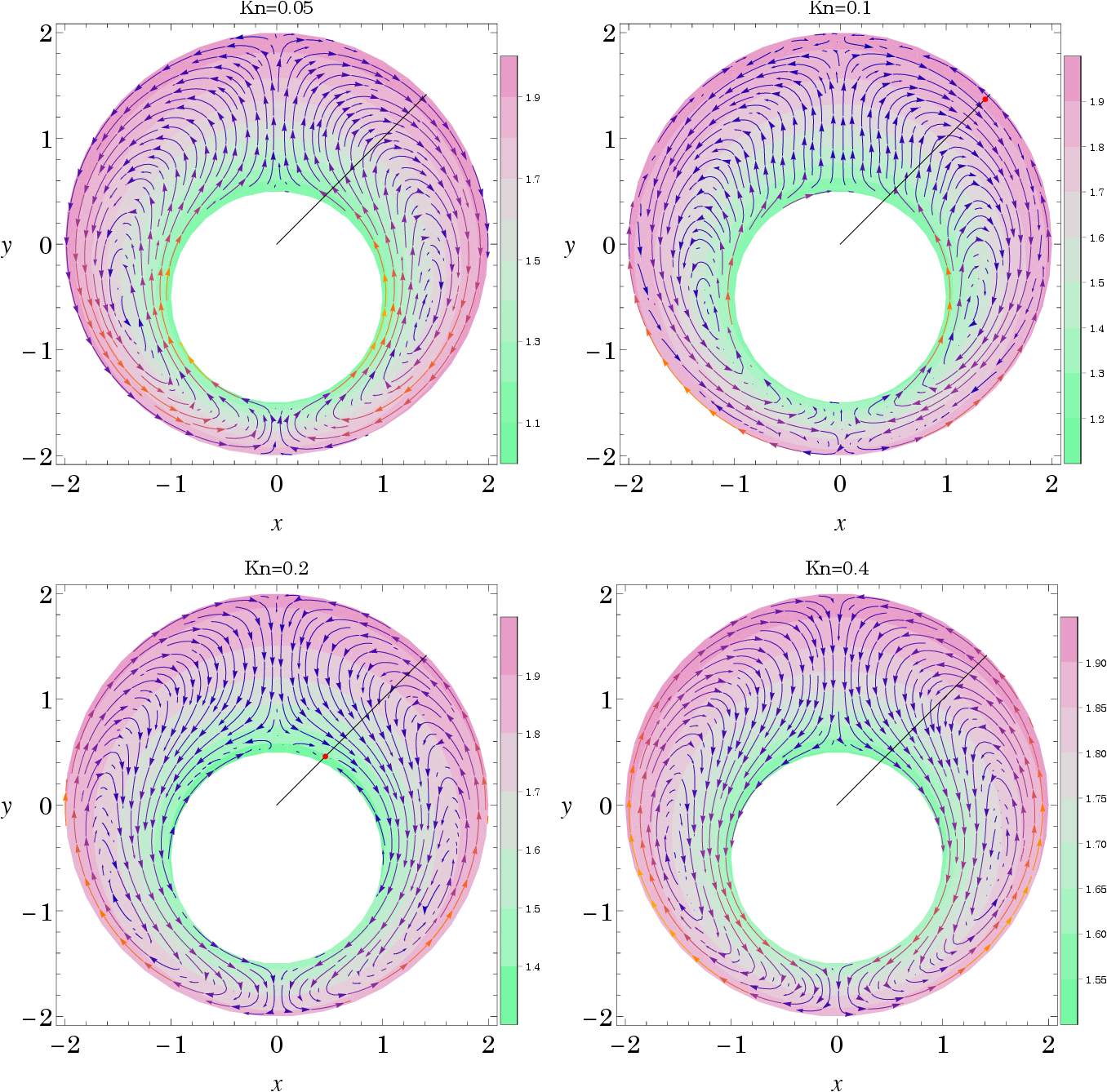}
    \caption{\label{fig:fem_mfs_stream}Velocity streamlines overlaid on temperature contours for different Knudsen numbers \(\mathrm{Kn} = 0.05, 0.1, 0.2, 0.4\) as predicted by the MFS.}
\end{figure}

In this problem, the gas flow is entirely driven by the temperature difference between the two cylinders without any external effect or gravity under consideration.
To gain insight into the velocity and temperature profiles, we visualize the velocity streamlines superimposed on temperature contours for different Knudsen numbers $\mathrm{Kn}=0.05,0.1,0.2$ and $0.4$ in Fig.~\ref{fig:fem_mfs_stream}, as predicted by the MFS.
The parameters for the MFS are fixed at $\alpha=1.5$ and $d=0.07$ for these computations.
These streamline plots reveal the intricate interplay between thermal stress and thermal transpiration effects, which arise due to the stress and heat flux evolution equations in the R13 model.
For small \(\mathrm{Kn} = 0.05\), two counter-rotating circulation zones emerge: one in the left half and the other in the right half of the annular region. As \(\mathrm{Kn}\) increases to \(0.1\), two additional vortices begin to form near the outer cylinder which indicate a shift in the flow structure.
With a further increase in the Knudsen number to \(\mathrm{Kn} = 0.2\), the newly formed vortices near the outer cylinder intensify, while the inner vortices diminishes in strength.
 For even larger \(\mathrm{Kn} = 0.4\), the small inner vortices disappear completely, restoring a two-vortex system similar to that at \(\mathrm{Kn} = 0.05\), but with the flow directions reversed.
This transformation in flow behavior highlights the competition between thermal stress and thermal transpiration effects, which govern rarefied gas flows under temperature gradients.

\begin{figure}[!t]
\centering
\includegraphics[scale=0.81]{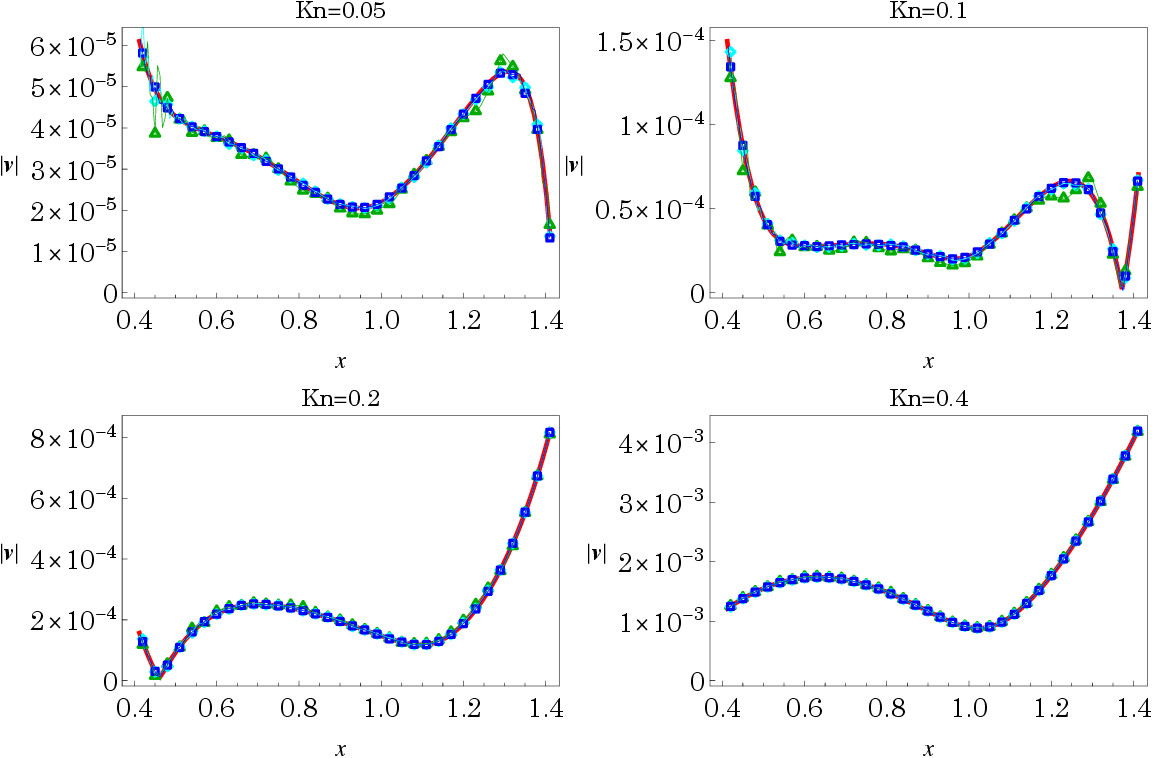} \vspace{3mm}
\includegraphics[scale=0.9]{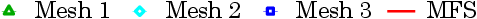}
\caption{\label{fig:fem_mfs_speed}Speed of the gas between the two cylinders along $y=x$ in the first quadrant for different Knudsen numbers.}
\end{figure}

To compare the results from MFS with those from FEM, we use three finest FEM meshes: Mesh 1 (\(\mathcal{T}_5\), coarsest), Mesh 2 (\(\mathcal{T}_6\), finer than Mesh 1), and Mesh 3 (\(\mathcal{T}_7\), finest).
Figure~\ref{fig:fem_mfs_speed} illustrates the speed of gas $|\bm{v}|$ along the line $y=x$ in the first quadrant  (or equivalently along $\vartheta=\pi/4$, as shown over Fig.~\ref{fig:fem_mfs_stream}) measured from the center of the outer cylinder for different Knudsen numbers $\mathrm{Kn}=0.05,0.1,0.2$ and $0.4$.
For small Knudsen numbers ($\mathrm{Kn}=0.05$ and $0.1$), the choice of FEM mesh significantly affects the results.
Meshes 1 and 2 are not refined enough to capture the gas speed accurately due to the small scale  ($\mathcal{O}(10^{-5})$), as shown by the green dashed (Mesh 1) and cyan dot-dashed (Mesh 2) lines in the top panels of Fig.~\ref{fig:fem_mfs_speed}.
For larger Knudsen numbers ($\mathrm{Kn}=0.2$ and $0.4$), the discrepancy between the three FEM meshes is significantly reduced and for $\mathrm{Kn}=0.4$, the results are nearly identical.
A reason for this behavior is that the magnitude of the velocity gets smaller with decreasing Knudsen number, which requires a finer mesh to resolve the flow features in the FEM.
In contrast, the MFS (solid red lines) exhibits stable convergence regardless of the Knudsen number or the grid spacing parameter.
The speed of the gas for $\mathrm{Kn}=0.1$ and $0.2$ is zero at $x=1.37$ and $x=0.46$, respectively.
These points correspond to the highlighted red dots in the streamline plot Fig.~\ref{fig:fem_mfs_stream}, at which the transition between the vortices along the inner and outer cylinder takes place for $\mathrm{Kn}=0.1$ and $0.2$.

Additionally, we calculate the heat flow rate through the inner cylinder defined as
\begin{align}
    Q_{\Gamma_1}=\int_{\Gamma_1} \bm{q}\cdot\bm{n} \, \text{d}l.
\end{align}
\begin{table}[t]
    \centering
    \renewcommand{\arraystretch}{1.2}
    \setlength{\tabcolsep}{10pt}

    \begin{tabular}{c|cc|cc|cc}
        \toprule
        \multicolumn{7}{c}{\textbf{FEM}} \\
        \midrule
        \textbf{Kn}
        & \multicolumn{2}{c|}{\textbf{Mesh 1}}
        & \multicolumn{2}{c|}{\textbf{Mesh 2}}
        & \multicolumn{2}{c}{\textbf{Mesh 3}} \\
        & $Q_{\Gamma_1}$ & Time
        & $Q_{\Gamma_1}$ & Time
        & $Q_{\Gamma_1}$ & Time \\
        \midrule
        0.05 & 1.5276481 & 5s & 1.5276241 & 29s & 1.5276212 & 185s \\
        0.1  & 2.4815209 & 5s & 2.4815120 & 30s & 2.4815119 & 185s \\
        0.2  & 3.5116585 & 5s & 3.5116914 & 29s & 3.5117014 & 187s \\
        0.4  & 4.1411597 & 5s & 4.1412806 & 30s & 4.1413130 & 191s \\
        \bottomrule
    \end{tabular}

    \vspace{1.2em}

    \begin{tabular}{c|cc|cc|cc}
        \toprule
        \multicolumn{7}{c}{\textbf{MFS}} \\
        \midrule
        \textbf{Kn}
        & \multicolumn{2}{c|}{$\bm{d = 0.15}\, (n_{\text{b}}=124)$}
        & \multicolumn{2}{c|}{$\bm{d = 0.1}\, (n_{\text{b}}=187)$}
        & \multicolumn{2}{c}{$\bm{d = 0.07}\, (n_{\text{b}}=268)$} \\
        & $Q_{\Gamma_1}$ & Time
        & $Q_{\Gamma_1}$ & Time
        & $Q_{\Gamma_1}$ & Time \\
        \midrule
        0.05 & 1.5276979 & 17s & 1.5276204 & 28s & 1.5276204 & 49s \\
        0.1  & 2.4815252 & 16s & 2.4815121 & 27s & 2.4815121 & 50s \\
        0.2  & 3.5117115 & 16s & 3.5117048 & 28s & 3.5117048 & 51s \\
        0.4  & 4.1413392 & 16s & 4.1413240 & 30s & 4.1413240 & 52s \\
        \bottomrule
    \end{tabular}

    \caption{Comparison of the heat flow rate through the inner cylinder $Q_{\Gamma_1}$ and computation time for FEM (top) and MFS (bottom) for different mesh refinements and source distances $d$ using 8 CPU cores.}\label{tab:comparison}
\end{table}
Table~\ref{tab:comparison} depicts the values of $Q_{\Gamma_1}$ obtained by considering different meshes for the FEM and different grid spacing (or number of boundary and singularity points) for different Knudsen numbers \(\mathrm{Kn} = 0.05, 0.1, 0.2, 0.4\).
We also calculate the time taken by the FEM and MFS to calculate $Q_{\Gamma_1}$ using different FEM meshes and grid spacing $d$ for the MFS using 8 CPU cores.
As the FEM mesh is refined from Mesh 1 to Mesh 3, the values of $Q_{\Gamma_1}$ converge, albeit with significantly increased computational time---reaching up to 191 seconds for Mesh 3.
However, the MFS achieves the accuracy up to $7$ significant digits with significantly lower computational cost.
For instance, in the finest FEM mesh (Mesh 3), the computation time reaches up to 191 seconds, while the most refined MFS case with $d = 0.07$ having $n_{\text{b}} = n_{\text{s}}= 268$ achieves a higher precision in less than a third of the time (approximately 52 seconds).
Additionally, even the coarser MFS configurations (e.g., with $d = 0.15$) yield accurate results with computation times as low as 16–17 seconds.
It has also been noticed that, when the grid spacing is reduced to $d = 0.07$, the MFS attains convergence in $Q_{\Gamma_1}$ values up to 10 decimal digits.
This highlights the MFS as not only a computationally efficient alternative to mesh-based solvers like FEM, but also a powerful method for achieving rapid convergence with high numerical accuracy in rarefied flow simulations.

\section{\label{sec:conclusion}Conclusion and Outlook}
In this work, we developed a generic approach of deriving the fundamental solutions for a linear moment system and their implementation in the MFS for studying rarefied gas flows, without the need to predefine Dirac-delta source terms.
The proposed approach has first been demonstrated for Stokes’ equations and then been extended to the R13 equations in two dimensions.
The derived fundamental solutions have been implemented successfully in the MFS solver and validated against analytical solutions to confirm their accuracy.
To further assess its performance, we have applied it to the problem of thermally-induced flow between two noncoaxial cylinders and compared the results with those from a FEM.
The MFS not only captured rarefaction effects accurately but also demonstrated computational efficiency.

A key advantage of the MFS lies in its mesh-free nature, which simplifies implementation and makes it particularly effective for the challenging thermally induced flows.
Additionally, we have investigated the impact of the placement of singularity points, grid spacing, and the effective condition number on the accuracy of the solution. Our findings highlight the importance of optimizing these parameters to balance the accuracy and numerical stability.
Overall, the generic MFS provides a robust and efficient alternative to traditional numerical methods for solving rarefied gas flow problems, with potential applications to other models in non-equilibrium flows.

The proposed MFS framework can be naturally extended to three-dimensional problems, which could be highly suitable for solving open-boundary configurations often encountered in microfluidic and aerospace applications.
Future directions include the treatment of inhomogeneous and nonlinear systems using iterative schemes such as the Picard iteration, as well as the extension to more complex moment models and kinetic-based closures.
Additionally, the methodology for deriving fundamental solutions of large linear systems could be leveraged in other boundary-type methods, such as the boundary element method (BEM), see, e.g.,~\cite{padrinoEfficientSimulationRarefied2022a}.

\section*{Acknowledgment}
Himanshi gratefully acknowledges the financial support from the Council of Scientific and Industrial Research (CSIR) [File No.:\ 09/1022(0111)/2020-EMR-I] and the Advanced Research Opportunities Program (AROP) offered by RWTH Aachen University. A.S.R.\ acknowledges the financial support from the Science and Engineering Research Board, India through the Grant No.\ MTR/2021/000417. L.T.\ and M.T.\ acknowledge the support by the German Research Foundation (DFG), project B04 (504291427) within the SFB 1481 (442047500).
\appendix
\numberwithin{equation}{section}
\section{\label{app:A}Analytic solution to the R13 equations}
To determine an analytic solution of the R13 equations, we substitute Eqs.~\eqref{R_bal} and~\eqref{m_bal} in Eqs.~\eqref{sigma_bal} and~\eqref{q_bal},  transforming the resulting system of equations~\eqref{mass_bal}--\eqref{q_bal} into the cylindrical coordinates $(r,\vartheta,z)$.
The choice of the cylindrical coordinates is natural, as the flow variables exhibit axial symmetry, making them invariant along the $z$-direction.
This approach has been previously employed to derive analytic solutions of the regularized 13-moment (R13) and regularized 26-moment (R26) equations in the linearized state for the problems of flow past a stationary cylinder or sphere~\citep{WT2012,Torrilhon2010,  RGST2021}.
The symmetry ansatz used in these studies assumes that the radial and angular dependencies of the variables can be separated, with angular dependencies being expressed using sine and cosine functions.
Specifically, the vector and tensor components having an odd number of indices in $\vartheta$ are selected to be proportional to $\sin \vartheta$ whereas the scalars and tensor components with an even number of indices in $\vartheta$ are made proportional to $\cos\vartheta$~\citep{Torrilhon2010}.
Furthermore, since the problem is quasi-two-dimensional, the dependency in the $z$-coordinate of the variables is automatically eliminated.
However, in the present problem, the rotation of the inner cylinder introduces an additional radial dependency. To account for this, extra functions dependent only on $r$ are included.
 Following the symmetry ansatz, the solution for the vectors $\bm{v}$ and $\bm{q}$ take the forms
\begin{align}
\label{vel_ansatz}
\bm{v}(r,\vartheta)=\begin{bmatrix}
a_0(r)+a(r)\,\cos{\vartheta}\\
b_0(r)-b(r)\,\sin{\vartheta}\\ 0
\end{bmatrix},\quad \text{and} \quad
\bm{q}(r,\vartheta)=\begin{bmatrix}
\alpha_0(r)+\alpha(r)\,\cos{\vartheta}\\
\beta_0(r)-\beta(r)\,\sin{\vartheta}\\ 0
\end{bmatrix},
\end{align}
that for the scalars $\theta$ and $p$ should take the form
\begin{align}
\label{p_ansatz}
\theta(r,\vartheta)=c_0(r)+c(r)\cos{\vartheta},\quad \text{and} \quad p(r,\vartheta)=d_0(r)+d(r)\cos{\vartheta},
\end{align}
and that for $\bm{\sigma}$ should take the form
\begin{align}
\label{stress_ansatz}
\bm{\sigma}(r,\vartheta)=\begin{bmatrix}
\gamma_0(r)+\gamma(r)\,\cos{\vartheta} & \kappa_0(r)+\kappa(r)\,\sin{\vartheta} & 0\\
\kappa_0(r)+\kappa(r)\,\sin{\vartheta} & -(\omega_0(r)+\omega(r)\,\cos{\vartheta}) & 0\\
0 & 0 & \sigma_{zz}
\end{bmatrix},
\end{align}
where $a_0(r)$, $a(r)$, $b_0(r)$, $b(r)$, $\alpha_0(r)$, $\alpha(r)$, $\beta_0(r)$, $\beta(r)$, $c_0(r)$, $c(r)$, $d_0(r)$, $d(r)$ $\gamma_0(r)$, $\gamma(r)$, $\kappa_0(r)$, $\kappa(r)$, $\omega_0(r)$ and $\omega(r)$ are the unknown functions that need to be determined, and $\sigma_{zz} = -\sigma_{rr}-\sigma_{\vartheta\vartheta}= -(\gamma_0(r) - \omega_0(r)+ \left(\gamma(r) - \omega(r)\right)\cos{\vartheta})$ as $\bm{\sigma}$ is a symmetric and trace-free tensor.
Insertion of ansatz~\eqref{vel_ansatz}--\eqref{stress_ansatz} in the R13 equations and separation of the radial and angular dependency leads to a system of 18 ordinary differential equations in the 18 unknowns.
The analytic solutions obtained using these ODEs consist of a bulk contribution—comprising logarithmic and polynomial terms in $r$ and $1/r$---and the Knudsen layer contributions, which involve modified Bessel functions of the first and second kinds.
The R13 equations predict three Knudsen layers, characterized by the eigenvalues $\lambda_1=\sqrt{5}/(3\mathrm{Kn}), \lambda_2=\sqrt{5}/(\sqrt{6}\mathrm{Kn})$ and $\lambda_3=\sqrt{3}/(\sqrt{2}\mathrm{Kn})$.
The bulk solution introduces twelve integration constants $c_i$ ($i=1,2,\dots, 12$) while the Knudsen layer part yields another twelve constants: $C_i^I\, (i=1,2,\dots,6)$ for the modified Bessel functions of the first kind and $C_i^K\, (i=1,2,\dots,6)$ for the modified Bessel functions of the second kind.
These constants are determined by enforcing boundary conditions at the inner and outer cylinders.
For brevity, explicit expressions for the obtained analytical solutions are not provided here.
\bibliographystyle{jfm_doi}
\bibliography{references}

\begin{thebibliography}{37}
\expandafter\ifx\csname natexlab\endcsname\relax\def\natexlab#1{#1}\fi
\def\au#1{#1} \def\ed#1{#1} \def\yr#1{#1}\def\at#1{#1}\def\jt#1{\textit{#1}}
  \def\bt#1{#1}\def\bvol#1{\textbf{#1}} \def\vol#1{#1} \def\pg#1{#1}
  \def\publ#1{#1}\def\arxiv#1{#1}\def\org#1{#1}\def\st#1{\textit{#1}}

\bibitem[Alves(2009)]{Alves2009}
{\sc \au{Alves, C.~J.}} \yr{2009}  \at{On the choice of source points in the
  method of fundamental solutions}.
  \href{http://dx.doi.org/10.1016/j.enganabound.2009.05.007}{ \jt{Eng. Anal.
  Bound. Elem.}} \href{http://dx.doi.org/10.1016/j.enganabound.2009.05.007}{
  \bvol{33},  \pg{1348--1361}}.

\bibitem[Banerjee(1994)]{Banerjee1994}
{\sc \au{Banerjee, P.}} \yr{1994} {\em The Boundary Element Methods in
  Engineering\/}.  \publ{McGraw-Hill}.

\bibitem[Berger \& Karageorghis(2001)]{BK2001}
{\sc \au{Berger, J.~R.} \& \au{Karageorghis, A.}} \yr{2001}  \at{The method of
  fundamental solutions for layered elastic materials}.
  \href{http://dx.doi.org/10.1016/S0955-7997(01)00002-9}{ \jt{Eng. Anal. Bound.
  Elem.}} \href{http://dx.doi.org/10.1016/S0955-7997(01)00002-9}{ \bvol{25},
  \pg{877--886}}.

\bibitem[Chen {\em et~al.\/}(2023)Chen, Noorizadegan, Young \& Chen]{CNYC2023}
{\sc \au{Chen, C.}, \au{Noorizadegan, A.}, \au{Young, D.} \& \au{Chen, C.-S.}}
  \yr{2023}  \at{On the determination of locating the source points of the
  {MFS} using effective condition number}.
  \href{http://dx.doi.org/10.1016/j.cam.2022.114955}{ \jt{J. Comput. Appl.
  Math.}} \href{http://dx.doi.org/10.1016/j.cam.2022.114955}{ \bvol{423},
  \pg{114955}}.

\bibitem[Chen {\em et~al.\/}(2016)Chen, Karageorghis \& Li]{CKL2016}
{\sc \au{Chen, C.~S.}, \au{Karageorghis, A.} \& \au{Li, Y.}} \yr{2016}  \at{On
  choosing the location of the sources in the {MFS}}.
  \href{http://dx.doi.org/10.1007/s11075-015-0036-0}{ \jt{Numer. Algorithms}}
  \href{http://dx.doi.org/10.1007/s11075-015-0036-0}{ \bvol{72},
  \pg{107--130}}.

\bibitem[Cheng {\em et~al.\/}(1994)Cheng, Antes \& Ortner]{CAO1994}
{\sc \au{Cheng, A.-D.}, \au{Antes, H.} \& \au{Ortner, N.}} \yr{1994}
  \at{Fundamental solutions of products of {H}elmholtz and polyharmonic
  operators}.
  \href{http://dx.doi.org/https://doi.org/10.1016/0955-7997(94)90095-7}{
  \jt{Eng. Anal. Bound. Elem.}}
  \href{http://dx.doi.org/https://doi.org/10.1016/0955-7997(94)90095-7}{
  \bvol{14},  \pg{187--191}}.

\bibitem[Cheng \& Hong(2020)]{CH2020}
{\sc \au{Cheng, A.~H.} \& \au{Hong, Y.}} \yr{2020}  \at{An overview of the
  method of fundamental solutions---{Solvability}, uniqueness, convergence, and
  stability}. \href{http://dx.doi.org/10.1016/j.enganabound.2020.08.013}{
  \jt{Eng. Anal. Bound. Elem.}}
  \href{http://dx.doi.org/10.1016/j.enganabound.2020.08.013}{ \bvol{120},
  \pg{118--152}}.

\bibitem[Claydon {\em et~al.\/}(2017)Claydon, Shrestha, Rana, Sprittles \&
  Lockerby]{CSRSL2017}
{\sc \au{Claydon, R.}, \au{Shrestha, A.}, \au{Rana, A.~S.}, \au{Sprittles,
  J.~E.} \& \au{Lockerby, D.~A.}} \yr{2017}  \at{Fundamental solutions to the
  regularised 13-moment equations: efficient computation of three-dimensional
  kinetic effects}. \href{http://dx.doi.org/10.1017/jfm.2017.763}{ \jt{J. Fluid
  Mech.}} \href{http://dx.doi.org/10.1017/jfm.2017.763}{ \bvol{833},  \pg{R4}}.

\bibitem[Donea \& Huerta(2003)]{doneaFiniteElementMethods2003}
{\sc \au{Donea, J.} \& \au{Huerta, A.}} \yr{2003} {\em Finite {{Element
  Methods}} for {{Flow Problems}}\/}, 1st edn.  \publ{Wiley}.

\bibitem[Drombosky {\em et~al.\/}(2009)Drombosky, Meyer \& Ling]{DML2009}
{\sc \au{Drombosky, T.~W.}, \au{Meyer, A.~L.} \& \au{Ling, L.}} \yr{2009}
  \at{Applicability of the method of fundamental solutions}.
  \href{http://dx.doi.org/10.1016/j.enganabound.2008.10.007}{ \jt{Eng. Anal.
  Bound. Elem.}} \href{http://dx.doi.org/10.1016/j.enganabound.2008.10.007}{
  \bvol{33},  \pg{637--643}}.

\bibitem[Fairweather {\em et~al.\/}(2003)Fairweather, Karageorghis \&
  Martin]{FKM2003}
{\sc \au{Fairweather, G.}, \au{Karageorghis, A.} \& \au{Martin, P.}} \yr{2003}
  \at{The method of fundamental solutions for scattering and radiation
  problems}. \href{http://dx.doi.org/10.1016/S0955-7997(03)00017-1}{ \jt{Eng.
  Anal. Bound. Elem.}} \href{http://dx.doi.org/10.1016/S0955-7997(03)00017-1}{
  \bvol{27},  \pg{759--769}}.

\bibitem[Folland(1995)]{Folland1995}
{\sc \au{Folland, G.~B.}} \yr{1995} {\em Introduction to Partial Differential
  Equations: Second Edition\/}, ned - new edition edn., ,  \vol{vol. 102}.
  \publ{Princeton University Press}.

\bibitem[Grad(1949)]{Grad1949}
{\sc \au{Grad, H.}} \yr{1949}  \at{On the kinetic theory of rarefied gases}.
  \href{http://dx.doi.org/10.1002/cpa.3160020403}{ \jt{Comm. Pure Appl. Math.}}
  \href{http://dx.doi.org/10.1002/cpa.3160020403}{ \bvol{2},  \pg{331--407}}.

\bibitem[Himanshi {\em et~al.\/}(2023)Himanshi, Rana \& Gupta]{HRG2023}
{\sc \au{Himanshi}, \au{Rana, A.~S.} \& \au{Gupta, V.~K.}} \yr{2023}
  \at{Fundamental solutions of an extended hydrodynamic model in two
  dimensions: {D}erivation, theory, and applications}.
  \href{http://dx.doi.org/10.1103/PhysRevE.108.015306}{ \jt{Phys. Rev. E}}
  \href{http://dx.doi.org/10.1103/PhysRevE.108.015306}{ \bvol{108},
  \pg{015306}}.

\bibitem[Himanshi {\em et~al.\/}(2025{\natexlab{{\em a\/}}})Himanshi, Rana \&
  Gupta]{HRG2025}
{\sc \au{Himanshi}, \au{Rana, A.~S.} \& \au{Gupta, V.~K.}}
  \yr{2025{\natexlab{{\em a\/}}}}  \at{Exploring external rarefied gas flows
  through the method of fundamental solutions}.
  \href{http://dx.doi.org/10.1103/PhysRevE.111.015101}{ \jt{Phys. Rev. E}}
  \href{http://dx.doi.org/10.1103/PhysRevE.111.015101}{ \bvol{111},
  \pg{015101}}.

\bibitem[Himanshi {\em et~al.\/}(2025{\natexlab{{\em b\/}}})Himanshi, Theisen,
  Rana, Torrilhon \& Gupta]{himanshi_2025_15279486}
{\sc \au{Himanshi}, \au{Theisen, L.}, \au{Rana, A.~S.}, \au{Torrilhon, M.} \&
  \au{Gupta, V.~K.}} \yr{2025{\natexlab{{\em b\/}}}} {R13\_MFS: A Meshless
  Solver for the Regularized 13-Moment Equations using the Method of
  Fundamental Solutions}.
  \href{https://doi.org/10.5281/zenodo.15279486}{Zenodo}.

\bibitem[Hörmander(1955)]{Hormander1955}
{\sc \au{Hörmander, L.}} \yr{1955}  \at{On the theory of general partial
  differential operators}. \href{http://dx.doi.org/10.1007/BF02392492}{
  \jt{Acta Mathematica}} \href{http://dx.doi.org/10.1007/BF02392492}{
  \bvol{94},  \pg{161--248}}.

\bibitem[Kupradze \& Aleksidze(1964)]{KA1964}
{\sc \au{Kupradze, V.~D.} \& \au{Aleksidze, M.~A.}} \yr{1964}  \at{The method
  of functional equations for the approximate solution of certain boundary
  value problems}. \href{http://dx.doi.org/10.1016/0041-5553(64)90006-0}{
  \jt{USSR Comput. Math. Math. Phys.}}
  \href{http://dx.doi.org/10.1016/0041-5553(64)90006-0}{ \bvol{4},
  \pg{82--126}}.

\bibitem[Lewintan {\em et~al.\/}(2025)Lewintan, Theisen \&
  Torrilhon]{lewintanWellPosednessR13Equations2025}
{\sc \au{Lewintan, P.}, \au{Theisen, L.} \& \au{Torrilhon, M.}} \yr{2025}
  Well-{{Posedness}} of the {{R13 Equations Using Tensor-Valued Korn
  Inequalities}},  \arxiv{arXiv: 2501.14108}.

\bibitem[Lockerby \& Collyer(2016)]{LC2016}
{\sc \au{Lockerby, D.~A.} \& \au{Collyer, B.}} \yr{2016}  \at{Fundamental
  solutions to moment equations for the simulation of microscale gas flows}.
  \href{http://dx.doi.org/10.1017/jfm.2016.606}{ \jt{J. Fluid Mech.}}
  \href{http://dx.doi.org/10.1017/jfm.2016.606}{ \bvol{806},  \pg{413--436}}.

\bibitem[Padrino {\em et~al.\/}(2022)Padrino, Sprittles \&
  Lockerby]{padrinoEfficientSimulationRarefied2022a}
{\sc \au{Padrino, J.~C.}, \au{Sprittles, J.~E.} \& \au{Lockerby, D.~A.}}
  \yr{2022}  \at{Efficient simulation of rarefied gas flow past a particle:
  {{A}} boundary element method for the linearized {{G13}} equations}.
  \href{http://dx.doi.org/10.1063/5.0091041}{ \jt{Physics of Fluids}}
  \href{http://dx.doi.org/10.1063/5.0091041}{ \bvol{34}~(6),  \pg{062011}}.

\bibitem[Rana {\em et~al.\/}(2013)Rana, Torrilhon \& Struchtrup]{RTS2013}
{\sc \au{Rana, A.}, \au{Torrilhon, M.} \& \au{Struchtrup, H.}} \yr{2013}  \at{A
  robust numerical method for the {R13} equations of rarefied gas dynamics:
  {A}pplication to lid driven cavity}.
  \href{http://dx.doi.org/10.1016/j.jcp.2012.11.023}{ \jt{J. Comput. Phys.}}
  \href{http://dx.doi.org/10.1016/j.jcp.2012.11.023}{ \bvol{236},
  \pg{169--186}}.

\bibitem[Rana {\em et~al.\/}(2021{\natexlab{{\em a\/}}})Rana, Gupta, Sprittles
  \& Torrilhon]{RGST2021}
{\sc \au{Rana, A.~S.}, \au{Gupta, V.~K.}, \au{Sprittles, J.~E.} \&
  \au{Torrilhon, M.}} \yr{2021{\natexlab{{\em a\/}}}}  \at{{$H$}-theorem and
  boundary conditions for the linear {R26} equations: application to flow past
  an evaporating droplet}. \href{http://dx.doi.org/10.1017/jfm.2021.622}{
  \jt{J. Fluid Mech.}} \href{http://dx.doi.org/10.1017/jfm.2021.622}{
  \bvol{924},  \pg{A16}}.

\bibitem[Rana {\em et~al.\/}(2018)Rana, Gupta \& Struchtrup]{RGS2018}
{\sc \au{Rana, A.~S.}, \au{Gupta, V.~K.} \& \au{Struchtrup, H.}} \yr{2018}
  \at{Coupled constitutive relations: a second law based higher-order closure
  for hydrodynamics}. \href{http://dx.doi.org/10.1098/rspa.2018.0323}{
  \jt{Proc. Roy. Soc. A}} \href{http://dx.doi.org/10.1098/rspa.2018.0323}{
  \bvol{474},  \pg{20180323}}.

\bibitem[Rana {\em et~al.\/}(2021{\natexlab{{\em b\/}}})Rana, Saini,
  Chakraborty, Lockerby \& Sprittles]{RSCLS2021}
{\sc \au{Rana, A.~S.}, \au{Saini, S.}, \au{Chakraborty, S.}, \au{Lockerby,
  D.~A.} \& \au{Sprittles, J.~E.}} \yr{2021{\natexlab{{\em b\/}}}}
  \at{Efficient simulation of non-classical liquid--vapour phase-transition
  flows: a method of fundamental solutions}.
  \href{http://dx.doi.org/10.1017/jfm.2021.405}{ \jt{J. Fluid Mech.}}
  \href{http://dx.doi.org/10.1017/jfm.2021.405}{ \bvol{919},  \pg{A35}}.

\bibitem[Struchtrup(2005)]{Struchtrup2005}
{\sc \au{Struchtrup, H.}} \yr{2005} {\em Macroscopic Transport Equations for
  Rarefied Gas Flows\/}.  \publ{Berlin: Springer}.

\bibitem[Struchtrup \& Torrilhon(2003)]{ST2003}
{\sc \au{Struchtrup, H.} \& \au{Torrilhon, M.}} \yr{2003}  \at{Regularization
  of {G}rad's 13 moment equations: {D}erivation and linear analysis}.
  \href{http://dx.doi.org/10.1063/1.1597472}{ \jt{Phys. Fluids}}
  \href{http://dx.doi.org/10.1063/1.1597472}{ \bvol{15},  \pg{2668--2680}}.

\bibitem[Theisen \& Torrilhon(2021)]{TT2021}
{\sc \au{Theisen, L.} \& \au{Torrilhon, M.}} \yr{2021}  \at{{fenicsR13}: A
  tensorial mixed finite element solver for the linear {R13} equations using
  the {FEniCS} computing platform}. \href{http://dx.doi.org/10.1145/3442378}{
  \jt{ACM Trans. Math. Softw.}} \href{http://dx.doi.org/10.1145/3442378}{
  \bvol{47}}.

\bibitem[Theisen \& Torrilhon(2025)]{theisenFenicsR13TensorialMixed2025}
{\sc \au{Theisen, L.} \& \au{Torrilhon, M.}} \yr{2025} {{fenicsR13}}: {{A
  Tensorial Mixed Finite Element Solver}} for the {{Linear R13 Equations
  Using}} the {{FEniCS Computing Platform}} (v1.5).
  \href{https://doi.org/10.5281/zenodo.14938147}{Zenodo}.

\bibitem[Torrilhon(2010)]{Torrilhon2010}
{\sc \au{Torrilhon, M.}} \yr{2010}  \at{Slow gas microflow past a sphere:
  {A}nalytical solution based on moment equations}.
  \href{http://dx.doi.org/10.1063/1.3453707}{ \jt{Phys. Fluids}}
  \href{http://dx.doi.org/10.1063/1.3453707}{ \bvol{22},  \pg{072001}}.

\bibitem[Torrilhon(2016)]{TorrilhonARFM}
{\sc \au{Torrilhon, M.}} \yr{2016}  \at{Modeling nonequilibrium gas flow based
  on moment equations}.
  \href{http://dx.doi.org/10.1146/annurev-fluid-122414-034259}{ \jt{Annu. Rev.
  Fluid Mech.}} \href{http://dx.doi.org/10.1146/annurev-fluid-122414-034259}{
  \bvol{48},  \pg{429--458}}.

\bibitem[Torrilhon \& Sarna(2017)]{TS2017}
{\sc \au{Torrilhon, M.} \& \au{Sarna, N.}} \yr{2017}  \at{Hierarchical
  {B}oltzmann simulations and model error estimation}.
  \href{http://dx.doi.org/https://doi.org/10.1016/j.jcp.2017.04.041}{ \jt{J.
  Comput. Phys.}}
  \href{http://dx.doi.org/https://doi.org/10.1016/j.jcp.2017.04.041}{
  \bvol{342},  \pg{66--84}}.

\bibitem[Wang {\em et~al.\/}(2018)Wang, Liu \& Qu]{WLQ2018}
{\sc \au{Wang, F.}, \au{Liu, C.-S.} \& \au{Qu, W.}} \yr{2018}  \at{Optimal
  sources in the {MFS} by minimizing a new merit function: Energy gap
  functional}. \href{http://dx.doi.org/10.1016/j.aml.2018.07.002}{ \jt{Appl.
  Math. Lett.}} \href{http://dx.doi.org/10.1016/j.aml.2018.07.002}{ \bvol{86},
  \pg{229--235}}.

\bibitem[Westerkamp \& Torrilhon(2012)]{WT2012}
{\sc \au{Westerkamp, A.} \& \au{Torrilhon, M.}} \yr{2012}  \at{Slow rarefied
  gas flow past a cylinder: {A}nalytical solution in comparison to the sphere}.
  \href{http://dx.doi.org/10.1063/1.4769505}{ \jt{AIP Conf. Proc.}}
  \href{http://dx.doi.org/10.1063/1.4769505}{ \bvol{1501},  \pg{207--214}}.

\bibitem[Westerkamp \& Torrilhon(2019)]{WT2019}
{\sc \au{Westerkamp, A.} \& \au{Torrilhon, M.}} \yr{2019}  \at{Finite element
  methods for the linear regularized 13-moment equations describing slow
  rarefied gas flows}.
  \href{http://dx.doi.org/https://doi.org/10.1016/j.jcp.2019.03.022}{ \jt{J.
  Comput. Phys.}}
  \href{http://dx.doi.org/https://doi.org/10.1016/j.jcp.2019.03.022}{
  \bvol{389},  \pg{1--21}}.

\bibitem[Wong \& Ling(2011)]{WL2011}
{\sc \au{Wong, K.~Y.} \& \au{Ling, L.}} \yr{2011}  \at{Optimality of the method
  of fundamental solutions}.
  \href{http://dx.doi.org/10.1016/j.enganabound.2010.06.002}{ \jt{Eng. Anal.
  Bound. Elem.}} \href{http://dx.doi.org/10.1016/j.enganabound.2010.06.002}{
  \bvol{35},  \pg{42--46}}.

\bibitem[Young {\em et~al.\/}(2006)Young, Jane, Fan, Murugesan \&
  Tsai]{YJFMT2006}
{\sc \au{Young, D.~L.}, \au{Jane, S.~J.}, \au{Fan, C.~M.}, \au{Murugesan, K.}
  \& \au{Tsai, C.~C.}} \yr{2006}  \at{The method of fundamental solutions for
  {2D} and {3D Stokes} problems}.
  \href{http://dx.doi.org/10.1016/j.jcp.2005.05.016}{ \jt{J. Comput. Phys.}}
  \href{http://dx.doi.org/10.1016/j.jcp.2005.05.016}{ \bvol{211},  \pg{1--8}}.

\end{thebibliography}
\end{document}